\documentclass[a4paper,11pt]{article}
\pdfoutput=1
\usepackage{jheppub}
\usepackage{amsmath,amsthm,amssymb,braket,color,verbatim,adjustbox,multirow,enumerate,textcomp,gensymb,subfigure,graphicx,diagbox,pifont,tabulary,booktabs,stackengine,epstopdf,dcolumn,makecell,listings,bbm,mathtools,autobreak,slashed,float,array,colordvi,xfrac}
\usepackage{hyperref}
\hypersetup{colorlinks,bookmarksopen,bookmarksnumbered,citecolor=[rgb]{0,0.35,0.5},linkcolor=[rgb]{0,0.35,0.5},urlcolor=[rgb]{0,0.35,0.5},pdfstartview=FitH,linktocpage}

\definecolor{eprintLinks}{rgb}{0.4,0.4,0.4}
\definecolor{journalLinks}{rgb}{0,0.35,0.5}
\newcommand{\MYhref}[3][blueLinks]{\href{#2}{\color{#1}{#3}}}

\def\g{\gamma}
\def\D{\mathfrak{d}}
\def\O{{\cal O}}
\def\v{{\cal V}}
\def\u{{\cal U}}
\def\w{{\cal W}}
\def\d{\textrm{d}}
\def\i{\textrm{i}}
\def\e{\textrm{e}}

\def\bfp{{\bf p}}
\def\bfk{{\bf k}}
\def\bfv{\mbox{\boldmath$\v$\unboldmath}}
\def\bfu{\mbox{\boldmath$\u$\unboldmath}}

\def\kn{\lvert\bfk\rvert}
\def\sdla{\langle\!\langle}
\def\sdra{\rangle\!\rangle}
\def\Bdla{\Bigl\langle\!\!\Bigl\langle}
\def\Bdra{\Bigr\rangle\!\!\Bigr\rangle}
\def\Ea{\overline{E}}
\def\pa{\overline{p}}
\def\ka{\overline{k}}
\def\va{\overline{v}}

\renewcommand{\thefootnote}{\fnsymbol{footnote}}
\title{\boldmath Lorentz and CPT breaking in gamma-ray burst neutrinos from string theory\footnote{Published as JHEP 03 (2023) 230, \url{https://doi.org/10.1007/JHEP03(2023)230}}}

\author[a]{Chengyi Li}\emailAdd{lichengyi@pku.edu.cn}
\affiliation[a]{School of Physics, Peking University, Beijing 100871, China}
\author[a,b,c]{and Bo-Qiang Ma\footnote{Author to whom any correspondence should be addressed.}}\emailAdd{mabq@pku.edu.cn}
\affiliation[b]{Center for High Energy Physics, Peking University, Beijing 100871, China}
\affiliation[c]{Collaborative Innovation Center of Quantum Matter, Beijing, China}

\abstract{
Previous studies on high-energy gamma-ray burst neutrinos from IceCube suggest a neutrino speed variation at the Lorentz violation~(LV) scale of $\sim 6.4\times 10^{17}$~GeV, with opposite velocity variances between neutrinos and antineutrinos. Within a space-time foam model, inspired by string theory, we develop an approach to describe the suggested neutrino/antineutrino propagation properties with both Lorentz invariance and CPT symmetry breaking. A threshold analysis on the bremsstrahlung of electron-positron pair~($\nu\rightarrow\nu ee^{+}$) for the superluminal~(anti)neutrino is performed. We find that, due to the energy violation caused by the quantum foam, such reaction may be restricted to occur at sufficient high energies and could even be kinematically forbidden. Constraints on neutrino LV from vacuum $ee^{+}$ pair emission are naturally avoided. Future experiments are appealed to test further the CPT violation of cosmic neutrinos and/or neutrino superluminality. 
}


\arxivnumber{2303.04765}

\begin{document}
\maketitle
\flushbottom

\renewcommand{\thefootnote}{\arabic{footnote}}
\setcounter{page}{2}


\section{Introduction}
\label{sect:intro}

Astrophysical neutrinos are ideal portals to reveal the tiny Lorentz invariance violation~(LV) as postulated by some quantum gravity~(QG) theories~\cite{Jacob:2006gn,Amelino-Camelia:2009imt,Amelino-Camelia:2015nqa}. The IceCube collaboration has reported the discovery of ultrahigh-energy neutrinos of extragalactic origin, including a couple of PeV events~\cite{IceCube:2013cdg,IceCube:2014jkq,IceCube:2016uab}. Recent studies~\cite{Amelino-Camelia:2016fuh,Amelino-Camelia:2016ohi,Huang:2018ham,Huang:2019etr,Huang:2022xto} of events in the~(near-)TeV--PeV range suggest a linearly energy dependent speed variation of neutrinos through their associations with gamma-ray bursts~(GRBs). Analyses lead to a Lorentz violation scale of $\sim 6\times10^{17}$~GeV, comparable with that determined from GRB photons~\cite{Shao:2009bv,Zhang:2014wpb,Xu:2016zxn,Liu:2018qrg,Zhu:2021pmw}. More intriguingly, it is proposed~\cite{Huang:2018ham} that either neutrinos or antineutrinos travel faster than the constant light speed $c$,\footnote{Henceforth natural units in which $c=\hslash=1$ are adopted.} whereas the other ones go slower than unity. This can be explained by the CPT-odd feature of the linear Lorentz violation~\cite{Huang:2018ham,Huang:2019etr,Huang:2022xto}, and leads further to the Charge--Parity--Time~(CPT) reversal symmetry breaking between neutrinos and antineutrinos, or a matter--antimatter asymmetry~\cite{Zhang:2018otj}. But it is also found that the attempt to interpret such phenomenological picture with field-theoretic models of LV faces challenges due to the constraints on the superluminal neutrino velocity and the corresponding LV from the kinematically allowed anomalous channels, e.g., vacuum pair emission~($\nu\rightarrow\nu ee^{+}$)~\cite{Zhang:2018otj}.

The main objective of the study we are here performing is to indicate that the experimental finding of LV for GRB neutrinos~\cite{Amelino-Camelia:2016fuh,Amelino-Camelia:2016ohi,Huang:2018ham,Huang:2019etr,Huang:2022xto} may coincide with the predictions from certain QG scheme that cannot be cast in an effective field theory~(EFT) description, i.e., the quantum~(Liouville inspired) space-time foam model from string/D-brane theories. In fact, the main idea has been outlined in a letter~\cite{Li:2022sgs}, and in this paper we provide a thorough account of the calculations and elaborate on more detailed results through in depth discussions. This framework has also been used previously in explaining light speed variation from analyzing flight times of GRB photons~\cite{Shao:2009bv,Zhang:2014wpb,Xu:2016zxn,Liu:2018qrg,Zhu:2021pmw} in a consistent way~\cite{Li:2021stm}.

The prototype idea of the quantum structure of space-time at a microscopic level---``space-time foam'' devised by Wheeler~\cite{Wheeler:1955stf}---arises from the uncertainties of quanta. For string/brane theory, such nontrivial foamy structures are provided by solitonic defects in some Liouville-string inspired models~\cite{Ellis:1999stm,Ellis:2004stm}, according to which our Universe lives on a~(compactified) D(irichlet)3-brane, roaming in a higher-dimensional bulk space, punctured by a population of D0-branes in type I/IIA strings~\cite{Ellis:1999stm,Ellis:2004stm,Ellis:2008stm}, as we will consider below~(or in IIB superstrings, of wrapped-up D-branes which are effectively pointlike~\cite{Li:2009tt}). The D-brane defects~(``D-particles'') appear to a braneworld observer as flashing-on-and-off vacuum structures when they cross the brane. Their interaction with open-string Standard-Model~(SM) excitations involving capture/splitting process and subsequent recoil reduces local Lorentz invariance. Such models, dubbed string/D-defect~(space-time) foam, go beyond the local EFT approach to QG with a variety of applications to study a number of phenomena, such as the so-induced vacuum refraction for photons~\cite{Mavromatos:2010pk} and fermions~\cite{Ellis:2000sfd}, origin of neutrino masses~\cite{Mavromatos:2010fmg} and mixing~\cite{Alexandre:2008sfo,Mavromatos:2009sfm,Mavromatos:2005sfm}, as well as string cosmologies associated with the dark sector of the Universe~\cite{Mavromatos:2010sco}. 

Our aim is to show that the suggested neutrino speed variation can be explained by means of the CPT-breaking aspects of such stringy QG models with linear Lorentz violations. Constraints implied by vacuum pair emission by the superluminal neutrino are addressed and found to be consistent with the findings of Refs.~\cite{Amelino-Camelia:2016ohi,Huang:2018ham,Huang:2019etr,Huang:2022xto} in such a string theoretic context. We also propose several viable ways on testing the CPT violation in the neutrino sector with future~(astrophysical) observations. 

The paper is organized as follows. In Sec.~\ref{sect:string}, we introduce within the framework of stringy space-time foam a scenario that admits CPT-violating neutrinos and compute the dispersion relation, velocity and traveling times. In Sect.~\ref{sect:stneut} we discuss the phenomenological implications of the results obtained by associating IceCube observations with GRBs on stringy QG. Section~\ref{sect:superd} is devoted to elaborating on the plausible mechanism permitting a stable propagation \textit{in vacuo} for the neutrino species against superluminal decays in the model as reported in~\cite{Li:2022sgs}. In particular, the ways to escape the threshold constraints are given. To conclude, a summary and discussion of our results is depicted in Sec.~\ref{sect:concl}. 

\section{Stringy D(efect)-foam}
\label{sect:string}

Consider the isotropic D-foam framework, as portrayed by the seminal works~\cite{Ellis:1999stm,Ellis:2004stm,Ellis:2008stm} in this area, the capture/splitting of a neutral open string such as a neutrino by a D-particle causes a recoil motion of the latter, described by a deformed stringy $\sigma$-model operator: 
\begin{equation}
\label{eq:rbo}
V\ni\int_{\partial\w}\d\tau\,\epsilon\,\u_{\ell}x^{0}\Theta_{\epsilon}(x^{0})\partial_{n}x^{\ell},
\end{equation}
where $\partial_{n}$ is the normal derivative on the boundary of the worldsheet $\partial\w$, $\Theta_{\epsilon\rightarrow 0^{+}}(t)=\frac{1}{2\pi\i}\int_{-\infty}^{\infty}\frac{\d q}{q-\i\epsilon}\e^{\i qt}$, and $\bfu$ is the spatial part of the recoil 4-velocity of the D-defect, $\u_{\ell}=\v_{\ell}(1-\bfv^{2})^{-1/2}\equiv\v_{\ell}\g_{\v}$, which, for heavy~(nonrelativistic) D-particles, reduces to the ordinary 3-velocity that can be identified as 
\begin{equation}
\label{eq:drv}
\bfv=M_{s}^{-1}g_{s}\Delta\bfk~\rightarrow~\v_{\Vert}\simeq M_{s}^{-1}g_{s}\lambda^{(\ell)}k_{\ell},
\end{equation}
following the~(logarithmic) conformal field theory methods~\cite{Kogan:1995lca,Ellis:1996lrt,Mavromatos:1998nz}. Above, $\Delta\bfk$ is the momentum transfer during a collision, $g_{s}\ll 1$ is the string coupling, and $M_{s}$ is the string scale. Suffixes $\Vert$ denote components along brane longitudinal dimensions, i.e., $\ell=1,2,3$, and $\lambda$ is the ratio of $\Delta k_{\ell}$ with respect to the incoming neutrino momentum, that is, $\lambda^{(\ell)}=\Delta k_{\ell}/k_{\ell}$, taken to be stochastic Gaussian~\cite{Mavromatos:2005sfm} with the moments: 
\begin{equation}
\label{eq:sfc}
\sdla\lambda^{(\ell)}\sdra=0,\quad\sdla\lambda^{(\ell)}\lambda^{(m)}\sdra=\D^{2}\delta^{\ell m}.
\end{equation}
The variances $\D^{(\ell)}=\sqrt{\sdla(\lambda^{(\ell)})^{2}\sdra}\neq 0$ may in general differ from each direction $\ell$, and $\sdla\cdot\!\cdot\!\cdot\sdra$ denotes an average over both~(i) statistical collections of D-particles and~(ii) quantum stringy fluctuations~\cite{Mavromatos:1998nz}, treated by resummation over worldsheet genera. 

Liouville dressing to the vertex operator~(\ref{eq:rbo})~\cite{Ellis:2004stm} then induces an off-diagonal distortion in target-space geometry $\hat{g}$, with $0\ell$ components~\cite{Kogan:1995lca,Ellis:1999stm}: 
\begin{equation}
\label{eq:flm}
g_{0\ell}(x^{0},\v_{\Vert})\sim\epsilon^{2}\v_{\ell}t\theta(t)\e^{-2\epsilon t}\sim\v_{\ell}.
\end{equation}
This results in a local background of Finsler type: $g_{\alpha\beta}=\eta_{\alpha\beta}+h_{\alpha\beta}$, $h_{0\ell}=\v_{\ell\Vert}^{A}\sigma_{A}$, where $\sigma_{A}$, $A=1,2,3$ are appropriate flavor matrices. This metric deformation~(\ref{eq:flm}) then affects the dispersion relation of neutrinos with mass $m_{\nu}$ via $k^{\alpha}k^{\beta}g_{\alpha\beta}(k)=-m_{\nu}^{2}$, yielding 
\begin{align}
\label{eq:emr}
E(\bfk)&=\bfk\cdot{\bfv}_{\Vert}\pm\kn\biggl(1+\frac{m_{\nu}^{2}}{\bfk^{2}}+\Bigl({\bfv}_{\Vert}\cdot\frac{\bfk}{\kn}\Bigr)^{2}\biggr)^{1/2}\nonumber\\
&\simeq{\cal E}_{M}+\frac{g_{s}}{M_{s}}\sum_{\ell}(k^{\ell})^{2}\lambda^{(\ell)}+\frac{{\cal E}_{M}}{2}\frac{g_{s}^{2}}{M_{s}^{2}}\sum_{\ell}(k^{\ell})^{2}(\lambda^{(\ell)})^{2},
\end{align}
where ${\cal E}_{M}=\pm\sqrt{k^{\ell}k_{\ell}+m_{\nu}^{2}}$ denotes the Minkowski energy with indefinite signature. The flavor structures have been omitted in the above expressions by taking account of Eq.~(\ref{eq:sfc}), i.e., $\sdla\lambda_{A}^{(\ell)}\lambda{}_{B}^{(m)}\sdra=\D^{2}\delta^{\ell m}\delta_{AB}$, since one needs to average~(\ref{eq:emr}) over the D-particle populations, that is 
\begin{align}
\label{eq:ane}
\sdla E\sdra&\eqqcolon\Ea(k)\simeq\Bdla\pm\,\bigl\lvert{\cal E}_{M}\bigr\rvert\Bigl(1+\frac{(\bfv\cdot\bfk)^{2}}{2\bfk^{2}}\Bigr)\Bdra\nonumber\\
&\simeq\pm\Bigl(1+\frac{g_{s}^{2}}{2M_{s}^{2}}\sideset{}{'}\sum_{\ell}\bigl(\D^{(\ell)}\bigr)^{2}k_{\ell}^{2}\Bigr)\bigl(\ka^{2}\!+m_{\nu}^{2}\bigr)^{1/2},
\end{align}
for $\ka\gg m_{\nu}$, where $\ka\coloneqq\sdla\kn\sdra$ is the averaged modulus of $\bfk$, and $\sum^{\prime}(\D^{2}k^{\ell}k_{\ell})$ represents $\sim\sdla\sum(\lambda k^{\ell}k_{\ell})\sdra$. 

On the other hand, the kinematics of the defect/neutrino scattering further yields~\cite{Mavromatos:2012ii}, 
\begin{align}
\label{eq:sec}
E_{\textrm{i}}&=E_{\textrm{f}}+M_{D}(\g_{\v}-1)+\delta v,\\
\label{eq:smc}
\bfk_{\textrm{i}}&=\bfk_{\textrm{f}}+\Delta\bfk=\bfk_{\textrm{f}}+M_{D}\g_{\v}\bfv_{\Vert},
\end{align}
where $(E,\bfk)_{\textrm{i/f}}$ is the 4-momentum for initial/final state, $\delta v$ denotes the fluctuation of brane vacuum energy during the scattering, with D-particle mass being $M_{D}=M_{s}/g_{s}$. Then, the explicit formula of $\overline{E}_{\textrm{f}}$ arises~(on average) from Eqs.~(\ref{eq:sec}) and~(\ref{eq:ane}), by noting $\overline{E}_{\textrm{i}}=\Ea$, as 
\begin{align}
\label{eq:fae}
\overline{E}_{\textrm{f}}(\overline{\bfk}_{\textrm{f}})&=\overline{E}_{\textrm{i}}(\overline{\bfk}_{\textrm{i}})-\Bdla\Bigl(\frac{1}{2}M_{D}\bfv_{\Vert}^{2}+{\cal O}(\v^{4})\Bigr)\Bdra\nonumber\\
&\simeq\overline{E}_{\textrm{i}}(\overline{\bfk}_{\textrm{f}})-\frac{g_{s}}{2M_{s}}\sideset{}{'}\sum_{\ell}\bigl(\D^{(\ell)}\bigr)^{2}(k_{\textrm{f}}^{\ell})^{2},
\end{align}
where the momentum conservation~[Eq.~(\ref{eq:smc})] is used, i.e., $\overline{\bfk}_{\textrm{f}}=\overline{\bfk}_{\textrm{i}}$ given that $\sdla\bfv_{\Vert}\sdra=0$, and $\sdla\delta v\sdra=0$ is assumed. Note that Eq.~(\ref{eq:fae}) reflects total combined effects of the D-foam on neutrino energy--momentum relations from both metric distortion and capture/splitting. 

Noticeably, antiparticles of spin-$\frac{1}{2}$ fermions can be regarded as ``holes'' with negative energies, one arrives at the \textit{effectively CPT-violating} dispersion relations for neutrino~($\nu$) and antineutrino~($\bar{\nu}$) \textit{in vacuo}~(or, for Majorana neutrino with different chirality): 
\begin{equation}
\label{eq:gdr}
\begin{split}
\Ea_{\nu}(\overline{\bfk})&\coloneqq\lvert\overline{E}_{\textrm{f}}^{(+)}(\overline{\bfk}; M_{D}, m_{\nu})\rvert,\\
\Ea_{\bar{\nu}}(\overline{\bfk})&\coloneqq\lvert\overline{E}_{\textrm{f}}^{(-)}(\overline{\bfk}; M_{D}, m_{\nu})\rvert,
\end{split}
\end{equation}
where $(+)/(-)$ denotes positive/negative part of Eq.~(\ref{eq:fae}) so that a physical particle always has energy $\Ea>0$. 

It is essential to understand that the difference between $\nu$'s and $\bar{\nu}$'s in Eq.~(\ref{eq:gdr}) follows from the Dirac's proposal of ``hole theory''. This applies only to fermions such as neutrinos, but not to bosons, as it is based upon the exclusion principle. In fact, as was elucidated in Refs.~\cite{Ellis:2004stm,Ellis:2008stm,Li:2009tt,Mavromatos:2010pk,Li:2021stm} and will be mentioned in the next section, neutral bosons like photons are subject to propagate subluminally, independent of their helicities, in this ``medium'' of~(stringy) vacuum defects. The reason for that may be traced back to the fact that the propagator of a fermion is similar to the square root of that of a vector boson. 

\subsection{Propagation velocities}

For the discussion of the GRB neutrinos here of interest, we consider an \textit{isotropic} foam, which further requires $\lambda^{(\ell)}=\lambda$ for all $\ell=1,2,3$. In such a case, the asymmetric dispersion relation~(\ref{eq:gdr}) for high-energy~(anti)neutrinos in this D-foam geometry reads 
\begin{align}
\label{eq:ndr}
\Ea_{\nu}(\overline{\bfk})&=\ka-\frac{g_{s}}{2M_{s}}\D_{\nu}^{2}\ka^{2}+\O(1/M_{D}^{2}),\\
\label{eq:adr}
\Ea_{\bar{\nu}}(\overline{\bfk})&=\ka+\frac{g_{s}}{2M_{s}}\D_{\nu}^{2}\ka^{2}+\O(1/M_{D}^{2}),
\end{align}
which reduces to our result in Ref.~\cite{Li:2022sgs} once higher-order corrections are negligible. Here, $\sdla\lambda^{2}\sdra=(\D^{(\ell)})^{2}\vert_{\ell=1,2,3}\equiv\D_{\nu}^{2}>0$ can be naturally up to $\O(1)$. The relativistic limit is used to omit the mass term, $m_{\nu}\simeq 0$. 

To get group velocities, one can certainly employ the relation $v=\partial E/\partial\kn$; it is possible, though, that such velocity law fails in QG, and that~\cite{Ellis:2008stm} the free propagation of particles may even disentangle from their dispersion relations.\footnote{For that case, both neutrinos and antineutrinos would be subluminal due to causality-respectful delays experienced by them as interacting with the foam of~\cite{Ellis:2008stm}; whereas there could be superluminal propagation in a type IIB stringy D-foam~\cite{Li:2009tt}.} Nevertheless, from a conservative point of view, we shall still assume Hamiltonian dynamics, and as such, the dispersion relation~(\ref{eq:ndr}) then yields a deformed \textit{subluminal} neutrino velocity as 
\begin{equation}
\label{eq:ngv}
\va_{\nu}\coloneqq\frac{\partial\Ea}{\partial\ka}=1-g_{s}\frac{\D_{\nu}^{2}\ka}{M_{s}}\simeq 1-\O\Bigl(g_{s}\frac{n_{D}\Ea}{M_{s}}\Bigr),
\end{equation}
where we substituted the lowest order dispersion $\ka\simeq\Ea$. The parameter $\D_{\nu}^{2}$ depends on the density of D-particles, $n_{D}$, which can be essentially arbitrary in the model, as is the~(stringy) quantum-gravity mass $M_{\textrm{sQG}}\coloneqq M_{s}/(\D^{2}g_{s})$. Though $n_{D}=n_{D}(t)$ could in general vary with the cosmological epochs~\cite{Mavromatos:2010pk}, say, it might evolve when the time elapses, we introduce a hypothetical \textit{uniform} foam situation, i.e., $n_{D}(t_{\textrm{late}})=n_{D}^{\star}\simeq\textrm{const.}$, at relatively late eras of the Universe~(for, e.g., redshifts $\lesssim 10$). 

Similarly, from Eq.~(\ref{eq:adr}), the velocity defect, i.e., $\delta_{v}^{D}\coloneqq\va-1$, for an antineutrino propagating in a quantum D0-brane foam becomes 
\begin{equation}
\label{eq:agv}
0<\delta_{v}^{D}=\frac{\ka}{M_{\textrm{sQG}}}\propto\frac{n_{D}^{\star}}{M_{s}}\Ea,
\end{equation}
which implies that antineutrinos are \textit{superluminal} particles. That is an important feature of our approach toward D-foam induced LV neutrino propagation, and is crucial for generating desired phenomenologies which will be discussed shortly. For our purpose, symmetric corrections of $\O(\D^{2})$ in particle and antiparticle sectors are assumed in the discussion, such that the amounts of CPT violation are the same for both $\nu$'s and $\bar{\nu}$'s.~(An asymmetry between neutrino and antineutrino sectors can be involved once there will be a need to reconcile with phenomenological constraints.) 

\subsection{Lag in travel times of neutrinos}

Before closing this section, we remark on the relation of the induced phase-space dependent metrics~(\ref{eq:flm}) with Finsler geometries, which, over past few years, are known to play a role in new physics as well as quantum gravity. The Finsler structure depends on both coordinates and momenta, as is precisely the situation encountered in~(\ref{eq:flm}). In some sense, the D-particle recoil may be viewed as an example of Finsler geometry in string theory. One may define appropriately the Finsler norm from the metric $\hat{g}(\v)$: $F=[g_{\alpha\beta}(y(\v))y^{\alpha}y^{\beta}]^{1/2}$~(here $y$'s denote ``velocities'' in Finslerian frameworks thereof~\cite{Zhu:2022blp}), and discuss geodesics in such space-times as in~\cite{Zhu:2022blp}. 

However as we have seen above, the effects of D-foam go beyond those encoded in a Finsler-like metric. Since one need to take account of the statistical effects of the quantum-fluctuating D-defects, this leads to a more general structure: stochastic Finsler geometry~\cite{Mavromatos:2010sco}. In fact, for the isotropic foam, the stochasticity, $\sdla\v_{\ell}\sdra=0$ of~(\ref{eq:sfc}), implies the \textit{restoration} of Lorentz symmetry \textit{on average} but with nontrivial~(higher-order) fluctuations expressed as correlators by $\sdla\v_{\ell}\v_{m}\sdra\propto\D^{2}\delta_{\ell m}$, $\D^{2}\neq 0$. The effects of recoil distortions are washed out \textit{statistically} and the metric will be like a Riemannian metric. So one may deal with particle propagation in an expanding space-time of the Friedmann--Robertson--Walker~(FRW) type as usual. The kinematical aspects of the recoil-brane scattering on matter also enter in the dispersion relation and velocity. The latter cannot be cast by a Finsler treatment. 

Based upon the above considerations, we estimate the flight time differences of neutrinos of astrophysical origin in a FRW Universe, where the momentum of the neutrino is redshifted by cosmological expansion. Given that the scale factor $a$ is related to the cosmic redshift $z^{\prime}$ by $a=\frac{1}{1+z^{\prime}}$, the velocity at the time of $z^{\prime}$ becomes 
\begin{equation}
\label{eq:rgv}
\va(\ka,z^{\prime})\simeq 1\mp g_{s}\D_{\nu}^{2}\Bigl(\frac{\ka/a}{M_{s}}\Bigr)=1\mp\D_{\nu}^{2}(1+z^{\prime})\Bigl(\frac{\ka}{M_{D}}\Bigr).
\end{equation}
The comoving distance $x_{c}$ between the source of the neutrinos, for example a GRB, and the Earth is 
\begin{equation}
\label{eq:cmd}
x_{c}=\int_{t_{0}}^{t_{\textrm{h/l}}}\va(\ka)\frac{\textrm{d}t}{a},
\end{equation}
where $t_{\textrm{h/l}}$ is the time when the high-/low-energy neutrinos arrive. If they are emitted at the same time $t_{0}$ at the GRB, the time lag can be obtained, 
\begin{equation}
\label{eq:td1}
\delta t\coloneqq t_{\textrm{h}}-t_{\textrm{l}}\simeq\pm\,\D_{\nu}^{2}\frac{\delta\ka}{M_{D}}\int_{0}^{z}\frac{(1+z^{\prime})}{H(z^{\prime})}\textrm{d}z^{\prime},
\end{equation}
where $\delta\ka=\ka_{\textrm{h}}-\ka_{\textrm{l}}$, and the Hubble parameter $H(z^{\prime})$ depends on the cosmological model. Since for late eras, at which we assume an approximately constant density of defects, the Hubble expansion on the D3-brane world is not affected~\cite{Mavromatos:2010sco}, the standard Cold-Dark Matter model with a positive cosmological constant~($\Lambda>0$) would be a good approximation of reality, as current observation suggests, then we have $H(z^{\prime})=H_{0}\sqrt{\Omega_{\textrm{M}}^{0}(1+z^{\prime})^{3}+\Omega_{\Lambda}^{0}}$, where $H_{0}$ is the Hubble constant. Hence, 
\begin{equation}
\label{eq:td2}
\delta t\simeq\pm\,\D_{\nu}^{2}H_{0}^{-1}\frac{\delta\Ea}{M_{D}}\int_{0}^{z}\textrm{d}z^{\prime}\frac{(1+z^{\prime})}{\sqrt{\Omega_{\textrm{M}}^{0}(1+z^{\prime})^{3}+\Omega_{\Lambda}^{0}}},
\end{equation}
here $\Ea\simeq\ka$ is identified as the neutrino~(antineutrino) energy observed on the Earth and $\delta\Ea\simeq E_{\textrm{h}}$. For neutrinos, there is a ``delay'' in their arrival times~($\delta t_{\nu}>0$), i.e., high-energy neutrinos arrive later compared to lower-energy ones; whereas for antineutrinos, $\delta t_{\bar{\nu}}<0$, corresponding to the arrival time ``advance''. 

\section{CPT-violating neutrino and phenomenological aspects}
\label{sect:stneut}

It has been realized, on the observational side, mainly over the last decade that high-energetic neutrinos are one of the most promising portals to LV physics~\cite{Jacob:2006gn,Amelino-Camelia:2009imt,Amelino-Camelia:2015nqa}. In this respect, Amelino-Camelia et al. developed a strategy of analysis on IceCube data to detect LV-modified laws of propagation for neutrinos in a quantum space-time~\cite{Amelino-Camelia:2016fuh}, and recently a Lorentz-violating picture is suggested by a series of model-independent studies on time-of-flight lags of IceCube neutrino events with respect to the purportedly associated light signals arriving from GRBs~\cite{Amelino-Camelia:2016ohi,Huang:2018ham,Huang:2019etr,Huang:2022xto}, which we proceed to review briefly. In the residue of this section we will show that this may serve as a support to the space-time foam model just presented. 

\subsection{Speed variation in IceCube GRB neutrinos}
\label{ssect:grb}

In Refs.~\cite{Amelino-Camelia:2016fuh,Amelino-Camelia:2016ohi,Huang:2018ham,Huang:2019etr,Huang:2022xto}, adopting a LV-deformed dispersion relation with a generic form, the authors get the modified propagation velocity for neutrinos as 
\begin{equation}
\label{eq:gpv}
v(E)=1-s_{n}\frac{1+n}{2}\Bigl(\frac{E}{E_{\textrm{LV},n}}\Bigr)^{n}
\end{equation}
where $n=1,2$ corresponds to linear or quadratic dependence of the energy, $s_{n}=\pm1$ is a sign factor of LV correction, and $E_{\textrm{LV},n}$ is the $n$th-order LV scale. Neutrino mass, which has been constrained within 1~eV~\cite{Huang:2014qwa}, can be safely neglected in the velocity formula for neutrinos with energies around or beyond TeV scale. Supposing the linear term~(i.e., $n=1$) in~(\ref{eq:gpv}) dominates, a regularity fitting well with TeV and PeV GRB neutrinos from IceCube~(including also  near-TeV events~\cite{Huang:2019etr}) is observed, indicating a \textit{neutrino speed variation} $v(E)=1-sE/E_{\textrm{LV}}^{\nu}$ with $s=\pm1$, at a scale of 
\begin{equation}
\label{eq:lvs}
E_{\textrm{LV}}^{\nu}=(6.4\pm1.5)\times 10^{17}~\textrm{GeV},
\end{equation}
which is close to the Planck scale $E_{\textrm{Pl}}\simeq 1.22\times 10^{19}$~GeV. Such an energy scale is consistent with various time-of-flight constraints, available today, from MeV neutrinos of supernova 1987A~\cite{Ellis:2008fc} and that from higher-energy neutrinos registered at IceCube, e.g.,~\cite{Wang:2016lne}; it is also compatible with the constraints~\cite{Ellis:2018ogq,Laha:2018hsh,Wei:2018ajw} from recent multimessenger observations of blazar TXS 0506+056 coincident with $\sim 290$ TeV track-like neutrino event~\cite{IceCube:2018dna}. We mention in passing that a light speed variation of $v(E)=1-E/E_{\textrm{LV}}^{\gamma}$ with $E_{\textrm{LV}}^{\gamma}\gtrsim 3.60\times 10^{17}$~GeV has also been suggested from time-of-flight studies on the energetic radiation from GRBs~\cite{Shao:2009bv,Zhang:2014wpb,Xu:2016zxn,Amelino-Camelia:2016ohi,Liu:2018qrg,Zhu:2021pmw} and from flares of active galactic nuclei~\cite{Li:2020uef}. This latter LV scale is not the same as the LV scale of neutrinos~(\ref{eq:lvs}) but of the same order of magnitude. 

Intriguingly, it is found that there exist both time ``delay''~($s=+1$) and ``advance''~($s=-1$) events, which can be interpreted with different propagation properties between neutrinos and antineutrinos~\cite{Huang:2018ham,Huang:2019etr,Huang:2022xto}, implying that neutrinos are superluminal and antineutrinos are subluminal, or vice versa, due to opposite signs of LV sign factor $s$. This proposal is at present only a hypothesis, but a reasonable one, because the IceCube detector cannot distinguish between neutrinos and antineutrinos~(except using the Glashow resonance, to be mentioned later). It is thus necessary to verify this hypothesis with further experimental tests, and to check then whether the revealed regularity~\cite{Amelino-Camelia:2016ohi,Huang:2018ham,Huang:2019etr,Huang:2022xto} can still persist or not with more neutrino events by IceCube~(or by other facilities) available in the future. 

\subsection{D-foam as an explanation for neutrino in-vacuo dispersion feature}
\label{ssect:interp}

Notice that the phenomenological LV picture revisited in the previous subsection is supported by a number of IceCube events among the gap over 4 orders of magnitude in neutrino energy scale, ranging from a few hundred GeV up to about 2~PeV, with all these GRB neutrino events falling on a pair of inclined lines surprisingly, see Refs.~\cite{Amelino-Camelia:2016ohi,Huang:2018ham,Huang:2019etr,Huang:2022xto} and also~\cite{Amelino-Camelia:2016fuh}. Such result yields a strong indication of a \textit{linear energy dependence} in the propagation speed of cosmic neutrinos, $\lvert v-1\rvert\sim\O(E/E_{\textrm{LV}}^{\nu})$, with $E_{\textrm{LV}}^{\nu}$, characterizing such a linear suppression of~(leading) LV effect, at about $10^{17}$~GeV, i.e., Eq.~(\ref{eq:lvs}). On the other hand, the presence of the opposite sign factors between neutrinos with their antiparticle counterparts~\cite{Huang:2018ham} clearly indicates a $\textit{CPT-violating propagation}$ of neutrinos in cosmic space. Hence, it is clear that the above two distinct LV properties, as exposed from current time-of-flight data~\cite{Huang:2018ham,Huang:2019etr,Huang:2022xto}, are consistent with the expected behavior for~(anti)neutrinos propagating in a gravitational ``medium'' of D-brane defects. 

Although it is not known from the analyses on IceCube data whether neutrinos, or antineutrinos, are superluminal, the D-particle foam scenario predicts that the latter would propagate slightly faster than the constant speed of light, so considering also the constraint that neutrinos and antineutrinos have the same amounts of speed variation, we can relate the D-foam induced velocity defect $\delta_{v}^{D}$ to the corresponding variation implied by the generic dispersion~(\ref{eq:gpv}). This yields, 
\begin{equation}
\label{eq:nvc}
\begin{split}
&\delta_{v}^{D}\simeq -\frac{E}{E_{\textrm{LV}}^{\nu}},\quad\textrm{for subluminal}~~\nu\textrm{'s};\\
&\delta_{v}^{D}\simeq\frac{E}{E_{\textrm{LV}}^{\nu}},\quad\textrm{for superluminal}~~\bar{\nu}\textrm{'s}.
\end{split}
\end{equation}

With the established correspondence~(\ref{eq:nvc}) for the neutrino velocities, we can then identify a combination of the foam parameters, $M_{D}/\D^{2}\simeq M_{\textrm{sQG}}$, i.e., stringy QG mass scale, as the linear-order Lorentz violation scale $E_{\textrm{LV}}$ determined by IceCube neutrinos: 
\begin{equation}
\label{eq:fpv}
\frac{\D_{\nu}^{2}g_{s}}{M_{s}}\simeq\frac{1}{E_{\textrm{LV}}^{\nu}}\simeq 1.6\times 10^{-18}~\textrm{GeV}^{-1},
\end{equation}
or, explicitly one has, 
\begin{equation}
\label{eq:qgv}
M_{\textrm{sQG}}^{(\nu)}\simeq E_{\textrm{LV}}^{\nu}\simeq 6.4\times 10^{17}~\textrm{GeV}.
\end{equation}
Such result gives an estimation of the neutrino QG scale $\sim 10^{17}$~GeV for the string/D-particle foam scenario. This is very much in line with one's intuitive expectation that the scale $M_{\textrm{sQG}}$ would be comparable to the Planck mass: $M_{\textrm{sQG}}\sim M_{\textrm{Pl}}~(\simeq 10^{19}$~GeV). 

Moreover, we note that the string model from the last section exhibits likewise nontrivial optical~($\gamma$) properties of the vacuum in terms of a velocity deformation for photon with different energy $\overline{\omega}$: 
\begin{equation}
\label{eq:pgv}
\delta_{v}^{D}=-\frac{\overline{\omega}}{M_{\textrm{sQG}}^{(\gamma)}}<0.
\end{equation}
This arises from the capture/splitting process of photon open strings by D-defects, in much the same way described in the previous section for the neutrino-D-particle interactions. The main difference is that photons are necessarily subluminal, thus, following our proposal~\cite{Li:2021stm} of interpreting the light speed variation from GRB photons with~(anisotropic) D-foam models, one can do the same thing in the context of isotropic foam scenario considered here. The arguments are similar to that of Ref.~\cite{Li:2021stm}~(see also~\cite{Li:2021tcv}), and will not be repeated here. It is essential to notice that, in that case, an estimate about the photon's QG scale can be got as $M_{\textrm{sQG}}^{(\gamma)}\simeq E_{\textrm{LV}}^{\gamma}\gtrsim 3.6\times 10^{17}$~GeV. Hence, we immediately observe 
\begin{equation}
\label{eq:qgr}
M_{\textrm{sQG}}^{(\nu)}\gtrsim M_{\textrm{sQG}}^{(\gamma)},
\end{equation}
i.e., $M_{\textrm{sQG}}$ for neutrinos may slightly be higher than that of photons but of the same order of magnitude. It should be noticed that this is exactly what the D-foam model expects since, from a theoretical viewpoint, the value of $M_{\textrm{sQG}}$ could depend upon quantities such as the string coupling $g_{s}$, the density of defects $n_{D}$ and, importantly, on the strength of particle interactions with foam defects, which may not be universal among particle species~\cite{Ellis:2003sfd}. 

Indeed, the neutrino interaction with the space-time D-defects could be very slightly suppressed compared with those of photons, because, in such a framework~\cite{Li:2022ugz}, only species that transform as trivial representations of the SM gauge group, i.e., neutral particles, are susceptible to the D-particle foam, in which case the fact that neutrinos are fermions with nontrivial $\textrm{SU}(2)_{L}$ properties renders such foam effects somewhat weakened. Nevertheless, neutrinos are at least electrically neutral excitations, so that the space-time foam appears not completely transparent to them, by invoking a possible background induced breaking of SU(2) gauge symmetry of the Standard Model. Hence the velocities of neutrinos would deviate from the low-energy velocity of light naturally less than photons, so that, $\lvert\delta_{v(\nu)}^{D}\rvert\lesssim\delta_{v(\gamma)}^{D}$, which just corresponds to the observation~(\ref{eq:qgr}) inferred from the findings of speed variation of high-energy cosmic neutrinos~\cite{Amelino-Camelia:2016ohi,Huang:2018ham,Huang:2019etr,Huang:2022xto} and cosmic photons~\cite{Shao:2009bv,Zhang:2014wpb,Xu:2016zxn,Liu:2018qrg,Zhu:2021pmw,Li:2020uef}. 

We should emphasize, in passing, that this avoids also the conflict, as indicated elsewhere~\cite{Crivellin:2020oov} for models where $\textrm{SU}(2)_{L}$ gauge invariance is still kept, with existing tight bounds on LV for the charged leptons~(see, e.g., Ref.~\cite{Li:2022ugz,He:2022jdl} and those quoted in~\cite{Crivellin:2020oov}). It is essential to notice that, in the D-foam case, Lorentz violation in neutrinos \textit{does not} imply LV in the charged lepton~(which is actually Lorentz invariant for specifically stringy reasons to be mentioned later), or vice versa. Therefore the result~(\ref{eq:fpv}) would \textit{not} lead to unacceptably large Lorentz-violating effects in the charged-lepton~(e.g., electron/positron) sectors, and the difficulty encountered by~\cite{Crivellin:2020oov} for the field-theoretic LV-scenario explanation does not arise when exploring a D-foam interpretation. 

From Eq.~(\ref{eq:fpv}), we can further deduce, 
\begin{equation}
\label{eq:svv}
\D_{\nu}^{2}\sim 10^{-18}\,\frac{M_{s}}{g_{s}}~\textrm{GeV}^{-1}.
\end{equation}
This means that the dimensionless variance $\D^{2}$, expressing stochastic fluctuations of the D-particle recoil velocity, relates to the value of the mass of the foam defect $M_{s}/g_{s}$, which in the modern version of string theory is essentially a free parameter. So for $\D^{2}\sim\O(1)$ for instance, a sub-Planckian D-particle mass $M_{s}/g_{s}\sim 10^{18}$~GeV is required so that the D-foam generates the phenomenologically suggested neutrino speed variation, by means of a Lorentz-/CPT-violating term $(g_{s}/M_{s})\D^{2}k^{2}$ between $\nu$ and $\bar{\nu}$ sectors. 

We should note that this type of CPT symmetry breaking may also lead to the observed baryon asymmetry in a natural way through gravitational leptogenesis of the early Universe~\cite{Mavromatos:2012ii}, which requires $M_{s}/g_{s}\sim 10^{25}\D^{2}~\textrm{GeV}$ so as to provide the physically observed lepton and, thus, baryon asymmetry. This seems to go beyond our result in Eq.~(\ref{eq:svv}). However, as mentioned above, in space-time foam situations, the density of D-particles $n_{D}$, which essentially the CPT-violating parameter, $\D^{2}$, as a~(Gaussian) variance of the fraction variable $\lambda$, depends upon, may vary with the cosmological time scale $t$, in the sense that their bulk distribution may not be constant all the time. We may expect, in general, a time dependent variance, $\sdla\lambda^{2}\sdra(t)\equiv\D^{2}(t)\propto n_{D}(t)$. Thus, it could be possible that at very early eras of the Universe there is a relatively \textit{dilute}, but still statistically significant, population of D-particles, by which a neutrino field can be encountered, thereby resulting in a somewhat small variance $\D$, upon averaging~(\ref{eq:drv}) over the D-particle populations. As the time elapses, the brane Universe, which roams in the bulk space, may however move into a region \textit{densely} populated with D-particles, so that for late epochs, say, redshifts $z\lesssim\O(10)$, neutrinos coming from cosmologically remote sources, either GRBs or active galaxies, interact with D-particles more frequently as they propagate over a distance in the foamy space-time, yielding stronger stochastic effects $\O(\D^{2})$. For a Grand-Unified-Theory scale $\sim\O(10^{15})$~GeV D-particle mass for instance, one acquires 
\begin{equation}
\label{eq:elv}
\D_{\nu}(t_{\textrm{early}})\simeq 10^{-5},~~\D_{\nu}(t_{\textrm{late}})\simeq 0.03,
\end{equation}
with corresponding distribution for the stochastic fluctuating  D-particle recoil velocities:\footnote{We stress here that the modeling of recoil operator $\v_{\ell}$~(or variable $\lambda$) by a Gaussian process is not based on analysis of the underlying string theory, but a reasonable assumption for characterizing stochasticity in recoiling space-time~(D-)defects~\cite{Mavromatos:2005sfm}.} 
\begin{equation}
\label{eq:gd}
{\cal D}(\lambda;\D^{2})\sim\exp\biggl[-\frac{1}{2}\biggl(\frac{\lambda}{\D(t)}\biggr)^{2}\biggr],
\end{equation}
which then differs with time scale $t$, particularly between early cosmological epochs, ${\cal D}_{\textrm{early}}$, and late eras, ${\cal D}_{\textrm{late}}$, for reasons explained above. Hence, in such sense, there is no conflict between our result~(\ref{eq:svv})~(or~(\ref{eq:qgv})) for explaining neutrino speed variation from GRBs with the constraint from CPT-violating effects in the early Universe~\cite{Mavromatos:2012ii}, and further, our estimate~(\ref{eq:elv}) indicates a much denser, but uniform to a large extent, bulk D-particle foam from late- to current-era Universe~($z\leq10$). 

In view of the above discussion, we conclude here that the finding of the neutrino speed variation of the form of $v(E)=1\mp E/E_{\textrm{LV}}^{\nu}$ from IceCube observations is compatible with the string/D-foam motivated deformation of velocity of neutrinos~(\ref{eq:ngv})~(and of antineutrinos~(\ref{eq:agv})) with $M_{\textrm{sQG}}\simeq 6\times 10^{17}$~GeV, approaching the Planck mass, through relating the stringy QG scale to the linear-order LV scale $E_{\textrm{LV}}^{\nu}$ from GRBs. It is crucial that the propagation properties suggested as differing between neutrinos and antineutrinos are explained by means of CPT-violating aspects of such D-brane foam model of Lorentz violation. This fact can thus in turn serve as a support to this type of models inspired by string theory. 

\section{Inhibited superluminal neutrino in-vacuo decay}
\label{sect:superd}

It was claimed however that assuming the usual conservation laws of energy and momentum, as widely postulated in field-theoretic LV models, superluminal neutrinos would dissipate much of their energies through kinematically allowed anomalous processes in a vacuum~\cite{Cohen:2011hx}, such as, Cherenkov radiation~($\nu\rightarrow\nu\gamma$), neutrino splitting~($\nu\rightarrow\nu\nu\bar{\nu}$), and bremsstrahlung of electron-positron pairs~($\nu\rightarrow\nu ee^{+}$), among which the last one, also known as vacuum $ee^{+}$ pair emission~(VPE), a neutral-current process, dominates the neutrino energy losing. Such processes would result in significant depletion of cosmic neutrino fluxes at high energies beyond which no neutrino should arrive at Earth. This has been used to bound superluminal velocities of high-energy neutrinos from IceCube~\cite{IceCube:2013cdg,IceCube:2014jkq,IceCube:2016uab,IceCube:2018dna}; in particular, powerful constraints on LV energy scale of $\gtrsim (10^{3}-10^{5})\times E_{\textrm{Pl}}$ are reported~\cite{Borriello:2013ala,Stecker:2013jfa,Diaz:2013wia,Wang:2020tej}. The same constraint applies to antineutrinos, if they are slightly superluminal, for the CP-conjugated processes as one can understand. 

However, these constraints become inapplicable, as we shall illustrate in the following, in the sense that the dominant energy-losing channels may be inhibited/forbidden in D-particle backgrounds, owing to so-induced deformation of energy--momentum conservation. 

To see this clearly, we perform here a threshold analysis on pair bremsstrahlung as an illustration. In the case of space-time D-foam, high-energy antineutrinos that are superluminal particles are expected to undergo such processes and to lose energy until they are at or below the threshold. Caution, however, is needed here. Due to the presence of recoiling D-defects near the braneworld, the energy conservation law in the common sense may be violated~\cite{Ellis:2000sf}, though, during reactions like $\bar{\nu}\rightarrow\bar{\nu} ee^{+}$, the total energy remains conserved once the kinetic energy of a heavy D-particle, $E_{D\textrm{-kin}}=M_{D}(\g_{\v}-1)\simeq\frac{1}{2}M_{D}\lvert\bfv\rvert^{2}$, is taken into account. Indeed, this fact has already been used in Eq.~(\ref{eq:sec}) as deriving the modified relativistic dispersion; in general, one has~\cite{Ellis:2000sf}, 
\begin{align}
\label{eq:ele}
\sum_{i}^{(N)}(\pm)E_{i}&\simeq\frac{1}{2}\frac{M_{s}}{g_{s}}\bfv_{(N)}^{2},\\
\label{eq:mle}
\sum_{i}^{(N)}(\pm)\bfp_{i}&\simeq M_{D}\bfv_{(N)},
\end{align}
for a multi-($N$-)particle reaction. Here, $\bfv_{(N)}$ is a proper generalization of~(\ref{eq:drv}), i.e., $\bfv=\sum_{i}(\pm)(\bfp_{i}/M_{D})$, and the notation $\sum(\pm)$ represents, for example in the case of $\bar{\nu}\rightarrow\bar{\nu}ee^{+}$ as we discuss now: 
\begin{align}
\label{eq:snd}
\sum_{i}(\pm)E_{i}&=E_{\textrm{in}}-\sum E_{\textrm{out}}\nonumber\\
&=E-(E^{\prime}+E_{e^{-}}+E_{e^{+}}),
\end{align}
where $E$, $E^{\prime}$ denote the energies for the incoming $\bar{\nu}$ and outgoing $\bar{\nu}$, respectively. Hence, the sum of the energies of all \textit{observable} particles is \textit{not} conserved due to a recoil-induced loss upon averaging over the foam: 
\begin{equation}
\label{eq:nel}
\sum_{i}^{(N)}(\pm)\Ea_{i}\eqqcolon\delta\Ea_{D}^{(N)}=\frac{M_{D}}{2}\sdla\bfv_{(N)}^{2}\sdra,
\end{equation}
which gives nonzero value due to the nontrivial stochastic fluctuations $\sdla\v^{\ell}\v_{\ell}\sdra\neq 0$, despite vanishing $\sdla\v_{\ell}\sdra$.

We should mention however that the energy lost $\delta\Ea_{D}$ due to stochastic interactions with the D-defects is relevant mainly for reactions with neutral elementary particles in the initial state, for which there is no obstacle to interact with the D-particles. In particular, the D-foam is predominantly transparent to~(electrically) charged matter on account of charge conservation~\cite{Ellis:2003sfd}, as justified by the strength of Cherenkov constraints for the electron/positron sectors~\cite{Li:2022ugz,He:2022jdl}.\footnote{The Crab Nebula severely limits electron LV effect with the tightest constraint: $E_{\textrm{LV}}^{e}\gtrsim 10^{26}~\textrm{GeV}\gg E_{\textrm{Pl}}$ for $v_{e}-1\simeq E/E_{\textrm{LV}}^{e}$~\cite{Li:2022ugz,He:2022jdl}, by using PeV $\gamma$-ray observed recently by LHAASO.} Hence, mainly the anomalous decay processes of the neutrino species, as discussed below, are affected dominantly by the foam effect. 

Consider now the superluminal $\bar{\nu}$ decay via $\bar{\nu}\rightarrow\bar{\nu}ee^{+}$, the averaged energy--momentum (non)conservation that follows from Eqs.~(\ref{eq:snd}) and~(\ref{eq:nel}) reads 
\begin{align}
\label{eq:aec}
\Ea&=\Ea{}^{\prime}+\Ea_{e^{-}}+\Ea_{e^{+}}+\delta\Ea_{D}^{(4)},\\
\label{eq:amc}
\ka&=\ka{}^{\prime}+\pa_{e^{-}}+\pa_{e^{+}},
\end{align}
where we denote the 4-momenta of the emitted $\bar{\nu}$ and $ee^{+}$ pairs as, $k^{\prime\alpha}=(\Ea{}^{\prime},\overline{\bfk}{}^{\prime})$, $p_{e/e^{+}}^{\alpha}=(\Ea,\overline{\bfp})_{e/e^{+}}$, respectively. The spatial 3-momentum is conserved on the average~(\ref{eq:amc}) due to the zero mean~(\ref{eq:sfc}) of $\bfv$ arising from isotropy. The Lorentz-violating energy defect, $\delta E$, is 
\begin{align}
\label{eq:lvd}
\Ea-\ka\eqqcolon\delta E(\overline{\bfk})\geq&~(\Ea_{e^{-}}-\pa_{e^{-}})\nonumber\\
&\,+(\Ea_{e^{+}}-\pa_{e^{+}})+\delta\Ea_{D}^{(4)}.
\end{align}
In D-particle foam models, Lorentz violation is absent for charged-lepton sectors, as explained:
\begin{equation}
\label{eq:edr}
\Ea_{e^{\mp}}=\sqrt{m_{e}^{2}+\pa_{e^{\mp}}^{2}}\simeq\pa_{e^{\mp}}\Bigl(1+\frac{m_{e}^{2}}{2\pa_{e^{\mp}}^{2}}\Bigr).
\end{equation}
Inserting the modified dispersion for~(anti)neutrinos~(\ref{eq:adr}) in Eq.~(\ref{eq:lvd}), and using~(\ref{eq:edr}) yields, 
\begin{equation}
\label{eq:the1}
\delta\Ea_{D}^{(4)}-\frac{g_{s}}{2M_{s}}\D_{\nu}^{2}\ka^{2}\leq -\frac{m_{e}^{2}}{2\pa_{e^{-}}}-\frac{m_{e}^{2}}{2\pa_{e^{+}}},
\end{equation}
from which the threshold condition can then be read off, by plugging in $E_{\textrm{thr}}\lesssim\Ea\simeq\ka=2\pa_{e/e^{+}}$: 
\begin{equation}
\label{eq:the2}
\Bigl(\delta\Ea_{D}^{(4)}-\frac{g_{s}}{2M_{s}}\D_{\nu}^{2}\Ea^{2}\Bigr)\Ea(\overline{\bfk})\simeq -2m_{e}^{2},
\end{equation}
where we used the fact that the $ee^{+}$ pair takes most of the total momentum, so that $E^{\prime}\simeq 0$ at threshold. Above, the energy violation, $\delta\Ea_{D}$, in a 4-particle interaction follows from Eq.~(\ref{eq:nel}) for $N=4$. The amount of such losses, during $\bar{\nu}$ Cherenkov-like decays in the D-foam backgrounds, can be estimated, to leading order, as 
\begin{equation}
\label{eq:sel}
\delta\Ea_{D}^{(4)}\simeq\frac{g_{s}}{M_{s}}2\varsigma_{I}^{\nu}\Ea^{2}+\ldots,
\end{equation}
where the dimensionless factor $\varsigma_{I}$ that controls the intensity of energy violation during reactions in which neutrinos~(antineutrinos) get involved is \textit{a priori} distinct from the QG parameter $\D^{2}$ that appears in the modified dispersions~(for details on~(\ref{eq:sel}), see Appendix~\ref{App:EnergyNonConservation} or~\cite{Ellis:2000sf}) but in principle of the same order of magnitude. The fact that $\delta\Ea_{D}\geq 0$~(or $\varsigma_{I}>0$), i.e., energy is \textit{lost} during particle interactions in the D-foam, follows from the underlying stringy treatment~\cite{Mavromatos:1998nz} of the recoil D-brane deformation. The condition that observed neutrinos are near or below the threshold energy then yields, 
\begin{equation}
\label{eq:thc}
-\frac{g_{s}}{2M_{s}}(4\varsigma_{I}^{\nu}-\D_{\nu}^{2})\ka^{3}\leq 2m_{e}^{2}.
\end{equation}

This inequality means that the stability of high-energy IceCube events can merely provide bounds on a \textit{combined quantity} of the fundamental parameters of the foam. For example, given the most energetic event, \#35 with energy of 2~PeV~\cite{IceCube:2013cdg},\footnote{A higher-energy $2.6$~PeV track event ATel~\#7856~\cite{IceCube:2016uab} is deduced to be a ``delay'' event~\cite{Huang:2018ham} which corresponds to a subluminal neutrino in our case, hence cannot be used to cast bounds from~(\ref{eq:thc}).} of the time ``advance'' type~\cite{Huang:2018ham}, and hence probably a superluminal antineutrino event, we then infer a limit from~(\ref{eq:thc}): 
\begin{equation}
\label{eq:fpc}
\lvert\D_{\nu}^{2}-4\varsigma_{I}^{\nu}\rvert g_{s}\sqrt{\alpha'}\lesssim 1.3\times 10^{-25}~\textrm{GeV}^{-1}.
\end{equation}
Here $\alpha^{\prime}$ is the string's Regge slope, related to the string mass scale via $M_{s}=(\alpha^{\prime})^{-1/2}$. 

It is obvious that the constraint~(\ref{eq:fpc}) set by the threshold effect of $\bar{\nu}\rightarrow\bar{\nu}ee^{+}$ for superluminal antineutrino with measured energy at $\sim 2$~PeV has nothing to do with the value we get for the stringy QG scale in Eq.~(\ref{eq:fpv})~(or~(\ref{eq:qgv})). To understand this situation from another viewpoint, we invoke the explicit form of VPE threshold in D-foam cases from the equation~(\ref{eq:thc}): 
\begin{equation}
\label{eq:thv}
E_{\textrm{thr}}^{\textrm{VPE}}\lesssim\biggl[\frac{M_{D}m_{e}^{2}}{(\D_{\nu}/2)^{2}-\varsigma_{I}^{\nu}}\biggr]^{1/3},\quad\textrm{for}~~\bar{\nu}\textrm{'s}.
\end{equation}
It now becomes clear that the above kinematical threshold is finite only if $4\varsigma_{I}<\D^{2}$; otherwise, it would be either infinite~(for $\D^{2}=4\varsigma_{I}$) or imaginary~($\D^{2}<4\varsigma_{I}$), implying that the pair emission is not permitted to happen in a vacuum. Even in the case of $\D^{2}>4\varsigma_{I}$, the process could be effectively inhibited if one has $\varsigma_{I}\approx(\D/2)^{2}$, in which case the energy threshold would be bumped up to a very high energy scale, for instance of order PeV, so that PeV-scale~(anti)neutrinos will not be depleted by $\bar{\nu}\rightarrow\bar{\nu}ee^{+}$. We are then able to observe superluminal events of such energy as detected by IceCube in spite of implying a very small difference between $\varsigma_{I}$ and $\frac{1}{4}\D^{2}$, namely, one may expect that~(in late-era Universe), 
\begin{align}
\label{eq:ipv}
\varsigma_{I}^{\nu}\sim 3\times 10^{-19}\,\frac{M_{s}}{g_{s}}~\textrm{GeV}^{-1},
\end{align}
and that, 
\begin{align}
\label{eq:cqc}
(\D_{\nu}^{2}-4\varsigma_{I}^{\nu})\lesssim 8.4\times 10^{-8},\ \ \textrm{for}~~\D_{\nu}^{2}>4\varsigma_{I}^{\nu},
\end{align}
which follows from~(\ref{eq:fpc}) and the fact that, for $\D^{2}\lesssim\O(1)$, we have $M_{s}/g_{s}<6.4\times 10^{17}~\textrm{GeV}$ from Eq.~(\ref{eq:fpv}). The constraint~(\ref{eq:cqc}), corresponding to a threshold energy around 2~PeV, which is at the desired order for observing PeV IceCube events, as explained, is \textit{not} unnaturally small in the context of D-particle foam models, where, as we have mentioned, both the parameters of the foam, $\D^{2}$ and $\varsigma_{I}$, are in general free to be adjusted. 

We emphasize the fact that there is energy loss in particle reactions, as a result of the nontrivial recoil of defects in the string/D-particle foam, distinguishes our approach from those field-theoretic models, where energy remains conserved in the usual sense though Lorentz invariance is broken. It is the energy losses caused by the quantum D-brane foam in this approach that raise the thresholds for superluminal neutrino decay, so as to permit a stable propagation. Hence, in such sense, very tight constraints so far cast by means of pair production threshold analyses~\cite{Borriello:2013ala,Stecker:2013jfa,Diaz:2013wia,Wang:2020tej}, which aim at limiting superluminal velocities within LV frameworks that entail a usual conservation of energy and momentum, are \textit{not} applicable to our case. This can be understood more clearly if inspecting again the foam-modified VPE threshold~(\ref{eq:thv}) but adopting the part of $M_{D}/\D^{2}$ from Eq.~(\ref{eq:fpv}): 
\begin{equation}
\label{eq:thv1}
E_{\textrm{thr}}^{\textrm{VPE}}\leq(4m_{e}^{2}{\cal E}_{*}^{\nu})^{1/3},
\end{equation}
or approximately, we may write $E_{\textrm{thr}}\simeq (4m_{e}^{2}{\cal E}_{*}^{\nu})^{1/3}$, corresponding to the emission of a zero energy antineutrino. Here a new energy scale ${\cal E}_{*}$ is defined as 
\begin{equation}
\label{eq:nes}
{\cal E}_{*}^{\nu}\coloneqq\frac{E_{\textrm{LV}}^{\nu}}{1-4\varsigma_{I}^{\nu}E_{\textrm{LV}}^{\nu}/M_{D}}.
\end{equation}
This indicates in a clear manner that the threshold limits as established in, e.g., Refs.~\cite{Borriello:2013ala,Stecker:2013jfa,Diaz:2013wia,Wang:2020tej}, are actually imposed on the scale ${\cal E}_{*}$, though can be evaded if ${\cal E}_{*}\leq 0$ or easily satisfied in case of ${\cal E}_{*}>0$ by proper assignment for the value of $\varsigma_{I}$, but \textit{not} imposed upon the \textit{actual} neutrino Lorentz violation scale $E_{\textrm{LV}}^{\nu}$, which corresponds to $M_{D}/\D^{2}$, as in Eq.~(\ref{eq:fpv}). The latter, as stressed, may only be limited by in-vacuo velocity dispersion effect via time-of-flight analyses, such as those performed in Refs.~\cite{Amelino-Camelia:2016fuh,Amelino-Camelia:2016ohi,Huang:2018ham,Huang:2019etr,Huang:2022xto}, wherein a neutrino speed variation at about $10^{17}$~GeV is suggested from IceCube neutrinos when associated with GRB candidates, or studies thereof~\cite{Wang:2016lne,Ellis:2018ogq,Laha:2018hsh,Wei:2018ajw}, where comparable flight-time bounds are obtained. Here, we end up with a consistent description for the findings of Refs.~\cite{Huang:2018ham,Huang:2019etr,Huang:2022xto} with the $\nu$ decay threshold constraint in the~(stringy) QG framework adopted here. 

We comment briefly here, for completeness, on a previous argument~\cite{Stecker:2014oxa} of a possible $\sim 2$~PeV cutoff in the $\nu(\bar{\nu})$ spectrum inferred from initial IceCube data set~\cite{IceCube:2013cdg}, which could be due to novel LV processes, particularly the neutrino splitting~(NS), which, as argued in~\cite{Stecker:2014oxa}, might play a more important role than pair emission by the neutrinos. Though the claim of this drop off and its interpretation with neutrino splittings are outdated, and actually disfavored by more recent data~(see below for details),\footnote{We are grateful to an anonymous referee who reminded us that the cutoff at about 2~PeV has proven an outdated proposal, and who encouraged us to comment on the role played by the newly observed Glashow resonance at IceCube~\cite{IceCube:2021rpz}.} it may still be useful to provide an additional refutation with our approach, by invoking the kinematical threshold for that LV splitting channel:\footnote{We postpone the discussion of superluminal $\bar{\nu}$-splitting, which is similar to that of vacuum pair emission, to Appendix~\ref{App:NeutrinoSplittingThreshold}.} 
\begin{equation}
\label{eq:sth}
E_{\textrm{thr}}^{\textrm{NS}}\simeq\biggl[9m_{\nu}^{2}\frac{M_{s}}{g_{s}(\D_{\nu}^{2}-2\varsigma_{I}^{\nu})}\biggr]^{1/3}.
\end{equation}
By adopting again the case in which $\D^{2}>4\varsigma_{I}(>2\varsigma_{I})$, and tuning their values, as was implemented in the VPE case, the anomalous decay through splitting into 3 neutrinos would then lead to a drop in $\bar{\nu}$~(not $\nu$) fluxes above 2~PeV. However, we find that this requires $(\D^{2}-2\varsigma_{I})<\O(10^{-18}\!-\!10^{-19})$, which is inconsistent with~(\ref{eq:cqc}). This implies that there exists some~(unnatural?) cancellation in $\D^{2}-2\varsigma_{I}$, indicating again, but from the point of view of the string model that the cutoff claimed elsewhere~\cite{Stecker:2014oxa} seems very suspicious. 

\section{Discussion and conclusion}
\label{sect:concl}

It is worthwhile to stress again that the finding of neutrino speed variation, and the consequent Lorentz/CPT violation for cosmic neutrinos as in Refs.~\cite{Amelino-Camelia:2016ohi,Huang:2018ham,Huang:2019etr,Huang:2022xto} and~\cite{Amelino-Camelia:2016fuh}, which we have devoted ourself to understand in Ref.~\cite{Li:2022sgs}, and in this paper, with a string theory model, still needs further comprehensive examination with data. As these results are of fundamental importance, it is necessary to check whether additional IceCube neutrino events can still support the revealed regularity~\cite{Amelino-Camelia:2016ohi,Huang:2018ham,Huang:2019etr,Huang:2022xto} or not, as well as checking the suggested subluminal and superluminal events, which, if indeed correspond, respectively, to neutrinos and antineutrinos, would be a strong sign for the type of CPT breaking as indicated by the theory. 

At present, as we mentioned previously, IceCube cannot distinguish between neutrinos from antineutrinos, so the intriguing correspondence uncovered by~\cite{Huang:2018ham,Huang:2019etr,Huang:2022xto} is only a conjecture from phenomenological point of view, while coincides with the prediction from a D-foam scenario, as explained thoroughly. There is an exception for electron antineutrinos at 6.3~PeV, at which the induced event rate is enhanced via the $W^{-}$ Glashow resonance channel~\cite{Glashow:1960zz}, as this phenomenon only occurs with $\bar{\nu}_{e}$'s. A candidate Glashow resonance event has lately been reported by IceCube~\cite{IceCube:2021rpz}, indicating the existence of $\gtrsim 6$~PeV cosmic antineutrinos. This may provide rare opportunities to test the result of~\cite{Huang:2018ham,Huang:2019etr,Huang:2022xto} and the D-foam models. Indeed, if one can infer through analysis according to the strategy of~\cite{Amelino-Camelia:2016fuh,Amelino-Camelia:2016ohi,Huang:2018ham} that such events correspond to ``advance''~(i.e., superluminal) events, it would largely favor the interpretation explored here, and would also allow one to tighten~(\ref{eq:cqc}) by a factor of at least $(6/2)^{3}\approx 27$. 

Together with the track-like event \#7856~\cite{IceCube:2016uab} with deposited energy of $\sim 2.6$~PeV~(which implies that the true neutrino energy can be of $\O(10)$~PeV), there appears to be no cutoff in the neutrino spectrum up to $\sim 10$~PeV---the highest energy for which current data are available. Still, this event~(ATel~\#7856) is inferred to be of the ``delay'' type according to prior analysis~\cite{Huang:2018ham} and hence probably a subluminal neutrino, which is stable in the quantum D-brane foam. It is then natural to observe such neutrino events from cosmologically remote sources, providing no extra constraint for the string model, as opposed to the case of antineutrinos. 

Thus, no clearly observable cutoff is produced once the effect of VPE and similar process is strongly inhibited for superluminal neutrino species with some pertinent choice of relevant QG parameters, as what current observation actually indicates. That fact might suggest stringy space-time foam as providing a possible realistic interpretation for all of these LV phenomenologies. 

We propose to search for higher-energy~(anti)neutrino events, which, if observed by IceCube or other neutrino telescopes in the future, would imply a much smaller difference between $\D^{2}$ and $4\varsigma_{I}$~(or, $2\varsigma_{I}$) for the superluminal picture. Of course, one is also encouraged to further examine the $\nu$- and $\bar{\nu}$-spectra, which, if both extend continuously~(or fall smoothly), without any cutoff, may indicate a more natural possibility that no splitting process or pair emission should ever happen for the antineutrino even if traveling \textit{in vacuo} slightly superluminally---a case can be naturally realized by $\varsigma_{I}\geq (\sfrac{1}{2})\D^{2}~(\Leftrightarrow\widetilde{\cal E}_{*}\leq 0)$ in the model. For details on our this argument, we refer the reader to Appendix~\ref{App:NeutrinoSplittingThreshold}. 

Before closing we mention that the stringent constraint provided by Lorentz violation deformed patterns in neutrino oscillations~\cite{IceCube:2014noc} can also be accounted for in D-foam situations, by recalling that the recoil and its resulting effects discussed above is essentially geometrical and kinematical, depending only upon neutrino energy~(and the status of CPT). As such, the D-brane foam is \textit{flavor blind}, with $M_{\textrm{sQG}}$ being the same for all neutrino species, hence with no effects on oscillations.\footnote{Note also that existing neutrino experiments~\cite{Lisi:2000ndc} are still far from achieving the required sensitivity~\cite{Alexandre:2008sfo} to detect QG decoherence effects on neutrino oscillations in D-foam backgrounds.} Thus the exotic contribution~(additive to that $L_{m}$ from the mass difference) to the flavor oscillation length, $L=2\pi/\lvert\Delta E\rvert$, like those arising in some generic quantum gravity models vanishes~(using $\D_{A}^{2}-\D_{B}^{2}=\Delta\D^{2}\sim 0$) in that case: 
\begin{equation}
\label{eq:qgol}
E\Delta\D_{\nu}^{2}L_{\textrm{sQG}}\sim 2\pi\frac{M_{D}}{E}\ \Rightarrow\ L_{\textrm{sQG}}\gg L_{m},
\end{equation}
and $L-L_{m}=\O(L_{m}/L_{\textrm{sQG}})$, from which it follows that $1/L\approx 1/L_{m}$, $L_{m}=4\pi E/\Delta m_{\nu}^{2}$. So, the present data~\cite{IceCube:2014noc} on neutrino oscillations do not prescribe the theory contemplated in Ref.~\cite{Li:2022sgs} and in this work with a universal scale $M_{\textrm{sQG}}\sim 10^{17}$~GeV. 

To summarize, a CPT-violating propagation of neutrinos~(antineutrinos), whose velocities scale linearly with their energies, can emerge from a string/D-brane theory model for space-time foam, being the type of \textit{isotropic}, Lorentz-invariant \textit{on average}, but Gaussian \textit{stochastically fluctuating}. Given that high-energy astrophysical neutrinos have gained a prominent role in probing such novel quantum-gravitational effect, we argue that the finding of the neutrino speed variation $v=1\mp E/E_{\textrm{LV}}^{\nu}$ at about $10^{17}$~GeV from IceCube GRB neutrino events~\cite{Huang:2018ham,Huang:2019etr,Huang:2022xto} can serve as a support for such theory. The existences of both time ``delay'' and ``advance'' events with equal amounts of speed variances are well accounted for by means of CPT-breaking aspects of such scenario, indicating that neutrinos are subluminal while antineutrinos travel at speeds in excess of the light speed $c$. For superluminal antineutrino, we show that the novel energy~(non)conservation implied by the recoiling D-particle foam may offer a viable mechanism in coping with the challenge due to the anomalous decay channels like electron-positron pair production. In this respect, those threshold constraints upon the superluminal velocities can be naturally avoided, and accordingly, being consistent with the findings of~\cite{Amelino-Camelia:2016ohi,Huang:2018ham,Huang:2019etr,Huang:2022xto} in this string-theory-inspired context. We appeal to further test both theoretically and experimentally the superluminality, Lorentz invariance violation and CPT violation for cosmic neutrinos. One may anticipate that more evidence will be observed as more neutrino data is accrued, enabling the stringy foam interpretation endorsed here to be either verified or falsified. Such efforts would contribute to fundamental physics as well as resolving important issues concerning the nature of space-time. 

\acknowledgments
This work was supported by the National Natural Science Foundation of China~(Grant No.~12075003). 

\appendix
\section{Energy violation in reactions}
\label{App:EnergyNonConservation}

We briefly review the issues of energy nonconservation during interactions in a recoiling D-foam. The treatment essentially follows~\cite{Ellis:2000sf}, of which the result is then applied to $\bar{\nu}$-decays by emitting $ee^{+}$ pairs for our purposes. 

Following the theorem highlighted in Ref.~\cite{Ellis:2000sf}, which states that during particle reactions the energy is in general violated, but being conserved for the \textit{complete system}~(i.e., matter plus recoiling D-particle inaccessible by a low-energy braneworld observer), i.e., Eq.~(\ref{eq:ele}), the energy violation in a simplest case, namely a $1\rightarrow 2$-body decay~(or a 3-particle vertex), is 
\begin{equation}
\label{eq:a1}
\Ea_{1}-\Ea_{2}-\Ea_{3}=\frac{1}{2}\frac{M_{s}}{g_{s}}\sdla\bfv_{(3)}^{2}\sdra,
\end{equation}
where $\bfv_{(3)}=[\bfp_{1}-(\bfp_{2}+\bfp_{3})]/M_{D}$ is the recoil velocity of a D-particle. We expect the average $\sdla\cdot\!\cdot\!\cdot\sdra$ over effects of the foam medium to be proportional to: 
\begin{equation}
\label{eq:a2}
\sdla\v^{2}\sdra\sim\O(p^{2})(\varsigma_{I}/M_{D}^{2}).
\end{equation}
Here $\varsigma_{I}$, as is made clear below, parametrizes energy loss during reactions, and may in general differ from the variance $\D^{2}$ appeared in particle dispersion relations. 

To estimate the above quantity, one may use the conservation for the~(average) 3-momenta, $\sdla M_{D}\bfv\sdra=0$, incorporating also the recoil of the background D-particle. This has been performed in~\cite{Ellis:2000sf}, where we refer the reader for details. Eq.~(\ref{eq:a1}) then becomes 
\begin{equation}
\label{eq:a3}
\delta\Ea_{D}^{(3)}=\frac{\varsigma_{I}}{M_{D}}\pa_{2}^{2}+\frac{\varsigma_{I}}{2M_{D}}\sum_{i}\delta p_{i}^{2}+\ldots,
\end{equation}
where $\pa_{2}$ is the momentum of a particle emitted during decay, say, particle 2, while particle 1 is converted into particle 3; $\delta p^{2}\equiv\sdla\bfp^2\sdra-\pa^2$ is minute variance of $\pa^2$, and the $\ldots$ denote stochastic quantum uncertainties. The latter two are purportedly negligible compared to $\pa_{2}^{2}/M_{D}$, hence can never cancel the leading term. 

We stress that, to leading order in $1/M_{D}\sim M_{\textrm{Pl}}^{-1}$, the matrix element for any particle reaction should in principle agree with that in low-energy relativistic field theory, while the ``recoil'' gravitational background that induces a violation of energy conservation would modify the usual \textit{kinematics} of quantum field-theoretic result for~(threshold) reactions. Multi-particle processes with $N\geq 4$ may be factorized as products of 3-point vertices and, $\delta\Ea_{D}$ is thus determined via summing up the corresponding violations in every 3-body interaction~(\ref{eq:a3}), irrelevant to the particle species considered. In this case, $\pa_{2}$ may denote a typical momentum of a mediator that may be exchanged in, e.g., a 4-particle interaction.

For the pair emission, i.e., $\bar{\nu}\rightarrow\bar{\nu}Z^{*}\rightarrow\bar{\nu}ee^{+}$, a 4-point amplitude which can be viewed as a product of two fundamental leptonic vertices mediated by the exchange of an off-shell $Z^{0}$ boson with momentum $\pa_{2}$, as in standard electroweak theory, the outgoing $\bar{\nu}$ with energy $\simeq 0$ at threshold, $\ka\simeq\pa_{2}=2\pa_{e}$, then yields, 
\begin{equation}
\label{eq:a4}
\delta\Ea_{D}^{(4)}\simeq\frac{2\varsigma_{I}^{\nu}}{M_{D}}\ka^{2}+\ldots\gtrsim\frac{g_{s}\varsigma_{I}^{\nu}}{M_{s}}2E_{\textrm{thr}}^{2},
\end{equation}
where $E_{\textrm{thr}}$ is the threshold for the vacuum pair emission, as given by~(\ref{eq:thv1}), and the foam parameter $\varsigma_{I}$ characterizes energy violation during processes involving the neutrino sector~(indicated by the superfix $\nu$).

\section{Neutrino splitting in D-foam}
\label{App:NeutrinoSplittingThreshold}

To understand how a superluminal antineutrino splits into three~($\bar{\nu}\rightarrow\bar{\nu}\nu\bar{\nu}$) in D-foam situations, it suffices to calculate its threshold energy. Let $\Ea, \ka$ be the~(average) energy and momentum of the incoming antineutrino, and $(\Ea{}^{\prime}, \overline{\bfk}{}^{\prime}), (\Ea, \overline{\bfk})_{\nu/\bar{\nu}}$ be the 4-momenta carried respectively by the outgoing $\bar{\nu}$, and by emitted $\nu\bar{\nu}$ pairs. The D-foam modified conservation law reads 
\begin{align}
\label{eq:b1}
\Ea&=\Ea{}^{\prime}+\Ea_{\nu}+\Ea_{\bar{\nu}}+\delta\Ea_{D}^{(4)},\nonumber\\
\ka&=\ka{}^{\prime}+\ka_{\nu}+\ka_{\bar{\nu}},
\end{align}
where the fact that all momenta are collinear at threshold is taken into account. For deformed dispersion relations with QG term $(\mp\sfrac{1}{2})(\D\ka)^{2}/M_{D}$, i.e., Eqs.~(\ref{eq:ndr}) and~(\ref{eq:adr}) but with mass-related term $\sdla m_{\nu}^{2}/(2\lvert\bfk\rvert)\sdra\simeq m_{\nu}^{2}/(2\ka)$, the threshold equation can be derived as 
\begin{align}
\label{eq:b2}
\frac{m_{\nu}^{2}}{\ka}&-\frac{m_{\nu}^{2}}{\ka{}^{\prime}}-\frac{m_{\nu}^{2}}{\ka_{\nu}}-\frac{m_{\nu}^{2}}{\ka_{\bar{\nu}}}\nonumber\\
&=2\times\delta\Ea_{D}^{(4)}+\frac{g_{s}\D_{\nu}^{2}}{M_{s}}\bigl((\ka{}^{\prime})^{2}-\ka^{2}-\ka_{\nu}^{2}+\ka_{\bar{\nu}}^{2}\bigr).
\end{align}
Note that the minute neutrino mass is \textit{not} negligible here, as this reaction exclusively involves the neutrino sector. By putting $\ka\simeq E_{\textrm{thr}}\simeq 3\ka{}^{\prime}=3\ka_{\nu/\bar{\nu}}$ therefore 
\begin{equation}
\label{eq:b3}
4m_{\nu}^{2}=-\ka\Bigl(\delta\Ea_{D}^{(4)}-\frac{g_{s}}{9M_{s}}4\D_{\nu}^{2}\ka^{2}\Bigr).
\end{equation}

The amount of energy violation, $\delta\Ea_{D}^{(4)}$, in this process, is again given by the sum of the violations in each of the two 3-point interactions~(\ref{eq:a3}), assuming factorization of relevant scattering amplitudes, via virtual $Z$ exchange, as $\bar{\nu}$-splitting can be viewed as a ``rotation'' of the neutrino-neutrino scattering process. Hence, 
\begin{equation}
\label{eq:b5}
\delta\Ea_{D}^{(4)}\simeq\frac{2\varsigma_{I}^{\nu}}{M_{D}}\Bigl(\frac{2\ka}{3}\Bigr)^{2}=\frac{8\varsigma_{I}^{\nu}}{9M_{D}}\ka^{2},
\end{equation}
where we assume that the three outgoing $\nu$~(or $\bar{\nu}$) each carry off approximately $1/3$ of the energy of the incoming $\bar{\nu}$, i.e., $\ka=\pa_{2}+\ka{}^{\prime}(=\ka/3)$, as used above. 

We end up with some remarks below based on the following kinematical threshold: 
\begin{equation}
\label{eq:b6}
E_{\textrm{thr}}^{\textrm{NS}}\simeq\sqrt[\uproot{20}3]{\frac{9M_{D}m_{\nu}^{2}}{(\D_{\nu}^{2}-2\varsigma_{I}^{\nu})}}=(9m_{\nu}^{2}\widetilde{\cal E}_{*}^{\nu})^{1/3},
\end{equation}
with
\begin{equation}
\label{eq:b7}
\widetilde{\cal E}_{*}^{\nu}\coloneqq\frac{E_{\textrm{LV}}^{\nu}}{1-2\varsigma_{I}^{\nu}E_{\textrm{LV}}^{\nu}/M_{D}}.
\end{equation}

The splitting process opens only if $\D^{2}>2\varsigma_{I}$, otherwise is forbidden, with the very case of $(\D,\varsigma_{I})=(0,0)$, yielding an infinite $E_{\textrm{thr}}$, as in the SM. If $\sqrt{2\varsigma_{I}}<\D\leq 2\sqrt{\varsigma_{I}}$, the decay channel through emitting $ee^{+}$ pairs is not allowed, though $\bar{\nu}$-splitting can still happen. This latter reaction can be suppressed at the kinematical level in case of $\varsigma_{I}\approx\frac{1}{2}\D^{2}$, thereby making the antineutrino practically stable \textit{in vacuo}, against any splitting effect at high energies. The threshold~(\ref{eq:b6}) is proportional to the mass of order $m_{\nu}\lesssim 1$~eV~\cite{Huang:2014qwa}. To push $E_{\textrm{thr}}$ to a sufficiently high scale, e.g., greater than 2~PeV, such that we are always below threshold when comparing with the IceCube data, an exceedingly small value of $(\D^{2}-2\varsigma_{I})$ would be required. A more natural solution to that could be $\varsigma_{I}\geq (\sfrac{1}{2})\D^{2}$, so that antineutrinos would never decay in D-foam geometries, despite traveling faster than light, and there is thus no need to fine-tune the model parameter.


\begin{thebibliography}{99}
\providecommand{\url}[1]{\texttt{#1}}
\providecommand{\urlprefix}{URL}
\providecommand{\eprint}[2][]{\url{#2}}

\bibitem{Jacob:2006gn}
U.~Jacob and T.~Piran, 
\MYhref[journalLinks]{https://doi.org/10.1038/nphys506}{Nat. Phys. {\bf 3}, 87 (2007)}.

\bibitem{Amelino-Camelia:2009imt}
G.~Amelino-Camelia and L.~Smolin, 
\MYhref[journalLinks]{https://doi.org/10.1103/PhysRevD.80.084017}{Phys. Rev. D {\bf 80}, 084017 (2009)}.

\bibitem{Amelino-Camelia:2015nqa}
G.~Amelino-Camelia, D.~Guetta and T.~Piran, 
\MYhref[journalLinks]{https://doi.org/10.1088/0004-637X/806/2/269}{ApJ {\bf 806}, 269 (2015)}.

\bibitem{IceCube:2013cdg}
IceCube Collaboration: M.~G.~Aartsen {et al.}, 
\MYhref[journalLinks]{https://doi.org/10.1103/PhysRevLett.111.021103}{Phys. Rev. Lett. {\bf 111}, 021103 (2013)};
\MYhref[journalLinks]{https://doi.org/10.1126/science.1242856}{Science {\bf 342}, 1242856 (2013)};
\MYhref[journalLinks]{https://doi.org/10.1103/PhysRevLett.113.101101}{Phys. Rev. Lett. {\bf 113}, 101101 (2014)}.

\bibitem{IceCube:2016uab}
IceCube Collaboration: M.~G.~Aartsen {et al.}, 
\MYhref[journalLinks]{https://doi.org/10.1103/PhysRevLett.117.241101}{Phys. Rev. Lett. {\bf 117}, 241101 (2016)}~[{\it Erratum} \MYhref[journalLinks]{https://doi.org/10.1103/PhysRevLett.119.259902}{{\bf 119}, 259902 (2017)}].

\bibitem{IceCube:2014jkq}
IceCube Collaboration: M.~G.~Aartsen {et al}., 
\MYhref[journalLinks]{https://doi.org/10.1088/2041-8205/805/1/L5}{ApJ Lett. {\bf 805}, L5 (2015)};
\MYhref[journalLinks]{https://doi.org/10.3847/0004-637X/824/2/115}{ApJ {\bf 824}, 115 (2016)};
\MYhref[journalLinks]{https://doi.org/10.3847/1538-4357/aa7569}{{\bf 843}, 112 (2017)}.

\bibitem{Amelino-Camelia:2016fuh}
G.~Amelino-Camelia, L.~Barcaroli, G.~D'Amico, N.~Loret and G.~Rosati, 
\MYhref[journalLinks]{https://doi.org/10.1016/j.physletb.2016.07.075}{Phys. Lett. B {\bf 761}, 318 (2016)}.

\bibitem{Amelino-Camelia:2016ohi}
G.~Amelino-Camelia, G.~D'Amico, G.~Rosati and N.~Loret, 
\MYhref[journalLinks]{https://doi.org/10.1038/s41550-017-0139}{Nat. Astron. {\bf 1}, 0139 (2017)}.

\bibitem{Huang:2018ham}
Y.~Huang and B.-Q.~Ma, 
\MYhref[journalLinks]{https://doi.org/10.1038/s42005-018-0061-0}{Commun. Phys. {\bf 1}, 62 (2018) doi:10.1038/s42005-018-0061-0}
[\MYhref[eprintLinks]{https://arxiv.org/abs/1810.01652}{{\ttfamily arXiv:1810.01652}}].

\bibitem{Huang:2019etr}
Y.~Huang, H.~Li and B.-Q.~Ma, 
\MYhref[journalLinks]{https://doi.org/10.1103/PhysRevD.99.123018}{Phys. Rev. D {\bf 99}, 123018 (2019)}.

\bibitem{Huang:2022xto}
Y.~Huang and B.-Q.~Ma, 
\MYhref[journalLinks]{https://doi.org/10.1016/j.fmre.2022.05.022}{Fundamental Research (2022) doi:10.1016/j.fmre.2022.05.022}.

\bibitem{Shao:2009bv}
L.~Shao, Z.~Xiao and B.-Q.~Ma, 
\MYhref[journalLinks]{https://doi.org/10.1016/j.astropartphys.2010.03.003}{Astropart. Phys. {\bf 33}, 312 (2010)}.

\bibitem{Zhang:2014wpb}
S.~Zhang and B.-Q.~Ma, 
\MYhref[journalLinks]{https://doi.org/10.1016/j.astropartphys.2014.04.008}{Astropart. Phys. {\bf 61}, 108 (2015)}.

\bibitem{Xu:2016zxn}
H.~Xu and B.-Q.~Ma, 
\MYhref[journalLinks]{https://doi.org/10.1016/j.astropartphys.2016.05.008}{Astropart. Phys. {\bf 82}, 72 (2016)};
\MYhref[journalLinks]{https://doi.org/10.1016/j.physletb.2016.07.044}{Phys. Lett. B {\bf 760}, 602 (2016)};
\MYhref[journalLinks]{https://doi.org/10.1088/1475-7516/2018/01/050}{JCAP {01} (2018) 050}.

\bibitem{Liu:2018qrg}
Y.~Liu and B.-Q.~Ma, 
\MYhref[journalLinks]{https://doi.org/10.1140/epjc/s10052-018-6294-y}{Eur. Phys. J. C {\bf 78}, 825 (2018)}.

\bibitem{Zhu:2021pmw}
J.~Zhu and B.-Q.~Ma, 
\MYhref[journalLinks]{https://doi.org/10.1016/j.physletb.2021.136518}{Phys. Lett. B {\bf 820}, 136518 (2021)};
\MYhref[journalLinks]{https://doi.org/10.1016/j.physletb.2021.136546}{{\bf 820}, 136546 (2021)}.

\bibitem{Zhang:2018otj}
X.~Zhang and B.-Q.~Ma, 
\MYhref[journalLinks]{https://doi.org/10.1103/PhysRevD.99.043013}{Phys. Rev. D {\bf 99}, 043013 (2019)}.

\bibitem{Li:2022sgs}
C.~Li and B.-Q.~Ma, 
\MYhref[journalLinks]{https://doi.org/10.1016/j.physletb.2022.137543}{Phys. Lett. B {\bf 835}, 137543 (2022)}.

\bibitem{Li:2021stm}
C.~Li and B.-Q.~Ma, 
\MYhref[journalLinks]{https://doi.org/10.1016/j.physletb.2021.136443}{Phys. Lett. B {\bf 819}, 136443 (2021)};
\MYhref[journalLinks]{https://doi.org/10.1016/j.rinp.2021.104380}{Results Phys. {\bf 26}, 104380 (2021)}.

\bibitem{Wheeler:1955stf}
J.~A.~Wheeler, 
\MYhref[journalLinks]{https://doi.org/10.1103/PhysRev.97.511}{Phys. Rev. {\bf 97}, 511 (1955)};
J.~A.~Wheeler, K.~Ford and M.~Goldhaber, 
\MYhref[journalLinks]{https://doi.org/10.1063/1.882666}{Phys. Today {\bf 52}, 63 (1998)}.

\bibitem{Ellis:1999stm}
J.~Ellis, N.~E.~Mavromatos and D.~V.~Nanopoulos, 
\MYhref[journalLinks]{https://doi.org/10.1023/A:1001852601248}{Gen. Rel. Grav. {\bf 32}, 127 (2000)};
\MYhref[journalLinks]{https://doi.org/10.1103/PhysRevD.61.027503}{Phys. Rev. D {\bf 61}, 027503 (1999)};
\MYhref[journalLinks]{https://doi.org/10.1103/PhysRevD.62.084019}{{\bf 62}, 084019 (2000)}.

\bibitem{Ellis:2004stm}
J.~Ellis, N.~E.~Mavromatos and M.~Westmuckett, 
\MYhref[journalLinks]{https://doi.org/10.1103/PhysRevD.70.044036}{Phys. Rev. D {\bf 70}, 044036 (2004)};
\MYhref[journalLinks]{https://doi.org/10.1103/PhysRevD.71.106006}{{\bf 71}, 106006 (2005)};
J.~Ellis, N.~E.~Mavromatos, D.~V.~Nanopoulos and M.~Westmuckett, 
\MYhref[journalLinks]{https://doi.org/10.1142/S0217751X06028990}{Int. J. Mod. Phys. A {\bf 21}, 1379 (2006)}.

\bibitem{Ellis:2008stm}
J.~Ellis, N.~E.~Mavromatos and D.~V.~Nanopoulos, 
\MYhref[journalLinks]{https://doi.org/10.1016/j.physletb.2008.06.029}{Phys. Lett. B {\bf 665}, 412 (2008)};
\MYhref[journalLinks]{https://doi.org/10.1016/j.physletb.2009.02.030}{{\bf 674}, 83 (2009)};
\MYhref[journalLinks]{https://doi.org/10.1016/j.physletb.2010.09.035}{{\bf 694}, 61 (2010)};
\MYhref[journalLinks]{https://doi.org/10.1142/S0217751X11053353}{Int. J. Mod. Phys. A {\bf 26}, 2243 (2011)}.

\bibitem{Li:2009tt}
T.~Li, N.~E.~Mavromatos, D.~V.~Nanopoulos and D.~Xie, 
\MYhref[journalLinks]{https://doi.org/10.1016/j.physletb.2009.07.062}{Phys. Lett. B {\bf 679}, 407 (2009)};
T.~Li and D.~V.~Nanopoulos,
\MYhref[journalLinks]{https://doi.org/10.1140/epjc/s10052-012-2044-8}{Eur. Phys. J. C {\bf 72}, 2044 (2012)}.

\bibitem{Mavromatos:2010pk}
N.~E.~Mavromatos, 
\MYhref[journalLinks]{https://doi.org/10.1142/S0217751X10050792}{Int. J. Mod. Phys. A {\bf 25}, 5409 (2010)}, and references therein.

\bibitem{Ellis:2000sfd}
J.~Ellis, N.~E.~Mavromatos, D.~V.~Nanopoulos and G.~Volkov, 
\MYhref[journalLinks]{https://doi.org/10.1023/A:1001980530113}{Gen. Rel. Grav. {\bf 32}, 1777 (2000)};
J.~Ellis, N.~E.~Mavromatos and D.~V.~Nanopoulos, 
\MYhref[journalLinks]{https://doi.org/10.1103/PhysRevD.65.064007}{Phys. Rev. D {\bf 65}, 064007 (2002)}.

\bibitem{Mavromatos:2010fmg}
N.~E.~Mavromatos, 
\MYhref[journalLinks]{https://doi.org/10.1103/PhysRevD.83.025018}{Phys. Rev. D {\bf 83}, 025018 (2011)};
J.~Ellis, N.~E.~Mavromatos and D.~V.~Nanopoulos, 
\MYhref[journalLinks]{https://doi.org/10.1103/PhysRevD.96.086012}{Phys. Rev. D {\bf 96}, 086012 (2017)}.

\bibitem{Alexandre:2008sfo}
J.~Ellis, N.~E.~Mavromatos and D.~V.~Nanopoulos, 
\MYhref[journalLinks]{https://doi.org/10.1103/PhysRevD.63.024024}{Phys. Rev. D {\bf 63}, 024024 (2000)};
J.~Alexandre, K.~Farakos, N.~E.~Mavromatos and P.~Pasipoularides, 
\MYhref[journalLinks]{https://doi.org/10.1103/PhysRevD.77.105001}{Phys. Rev. D {\bf 77}, 105001 (2008)};
\MYhref[journalLinks]{https://doi.org/10.1103/PhysRevD.79.107701}{{\bf 79}, 107701 (2009)}.

\bibitem{Mavromatos:2009sfm}
N.~E.~Mavromatos, S.~Sarkar and W.~Tarantino, 
\MYhref[journalLinks]{https://doi.org/10.1103/PhysRevD.80.084046}{Phys. Rev. D {\bf 80}, 084046 (2009)};
\MYhref[journalLinks]{https://doi.org/10.1142/S0217732313500454}{Mod. Phys. Lett. A {\bf 28}, 1350045 (2013)}.

\bibitem{Mavromatos:2005sfm}
N.~E.~Mavromatos and S.~Sarkar, 
\MYhref[journalLinks]{https://doi.org/10.1103/PhysRevD.72.065016}{Phys. Rev. D {\bf 72}, 065016 (2005)};
J.~Bernabeu, N.~E.~Mavromatos and S.~Sarkar, 
\MYhref[journalLinks]{https://doi.org/10.1103/PhysRevD.74.045014}{Phys. Rev. D {\bf 74}, 045014 (2006)}.

\bibitem{Mavromatos:2010sco}
N.~E.~Mavromatos, S.~Sarkar and A.~Vergou, 
\MYhref[journalLinks]{https://doi.org/10.1016/j.physletb.2010.12.045}{Phys. Lett. B {\bf 696}, 300 (2011)};
N.~E.~Mavromatos, V.~A.~Mitsou, S.~Sarkar and A.~Vergou, 
\MYhref[journalLinks]{https://doi.org/10.1140/epjc/s10052-012-1956-7}{Eur. Phys. J. C {\bf 72}, 1956 (2012)};
T.~Elghozi, N.~E.~Mavromatos, M.~Sakellariadou and M.~F.~Yusaf, 
\MYhref[journalLinks]{https://doi.org/10.1088/1475-7516/2016/02/060}{JCAP {02} (2016) 060}.

\bibitem{Kogan:1995lca}
I.~I.~Kogan and N.~E.~Mavromatos, 
\MYhref[journalLinks]{https://doi.org/10.1016/0370-2693(96)00195-5}{Phys. Lett. B {\bf 375}, 111 (1996)};
I.~I.~Kogan, N.~E.~Mavromatos and J.~F.~Wheater, 
\MYhref[journalLinks]{https://doi.org/10.1016/0370-2693(96)01067-2}{Phys. Lett. B {\bf 387}, 483 (1996)}.

\bibitem{Ellis:1996lrt}
J.~Ellis, N.~E.~Mavromatos and D.~V.~Nanopoulos, 
\MYhref[journalLinks]{https://doi.org/10.1142/S0217751X97001481}{Int. J. Mod. Phys. A {\bf 12}, 2639 (1997)};
\MYhref[journalLinks]{https://doi.org/10.1142/S0217751X98000470}{{\bf13}, 1059 (1998)}.

\bibitem{Mavromatos:1998nz}
N.~E.~Mavromatos and R.~J.~Szabo, 
\MYhref[journalLinks]{https://doi.org/10.1103/PhysRevD.59.104018}{Phys. Rev. D {\bf 59}, 104018 (1999)}.

\bibitem{Mavromatos:2012ii}
N.~E.~Mavromatos and S.~Sarkar, 
\MYhref[journalLinks]{https://doi.org/10.1140/epjc/s10052-013-2359-0}{Eur. Phys. J. C {\bf 73}, 2359 (2013)}.

\bibitem{Zhu:2022blp}
See, e.g., J.~Zhu and B.-Q.~Ma, 
\MYhref[journalLinks]{https://doi.org/10.1103/PhysRevD.105.124069}{Phys. Rev. D {\bf 105}, 124069 (2022)}, and references therein.

\bibitem{Huang:2014qwa}
Y.~Huang and B.-Q.~Ma, 
{The Universe {\bf 2}, 65--71 (2014)}
[\MYhref[eprintLinks]{https://arxiv.org/abs/1407.4357}{{\ttfamily arXiv:1407.4357}}].

\bibitem{Ellis:2008fc}
J.~Ellis, N.~Harries, A.~Meregaglia, A.~Rubbia and A.~S.~Sakharov, 
\MYhref[journalLinks]{https://doi.org/10.1103/PhysRevD.78.033013}{Phys. Rev. D {\bf 78}, 033013 (2008)}.

\bibitem{Wang:2016lne}
Z.-Y.~Wang, R.-Y.~Liu and X.-Y.~Wang, 
\MYhref[journalLinks]{https://doi.org/10.1103/PhysRevLett.116.151101}{Phys. Rev. Lett. {\bf 116}, 151101 (2016)}.

\bibitem{Ellis:2018ogq}
J.~Ellis, N.~E.~Mavromatos, A.~S.~Sakharov and E.~K.~Sarkisyan-Grinbaum, 
\MYhref[journalLinks]{https://doi.org/10.1016/j.physletb.2018.11.062}{Phys. Lett. B {\bf 789}, 352 (2019)}.

\bibitem{Laha:2018hsh}
R.~Laha, 
\MYhref[journalLinks]{https://doi.org/10.1103/PhysRevD.100.103002}{Phys. Rev. D {\bf 100}, 103002 (2019)}.

\bibitem{Wei:2018ajw}
J.-J.~Wei {et al}., 
\MYhref[journalLinks]{https://doi.org/10.1016/j.jheap.2019.01.002}{JHEAp {22} (2019) 1}.

\bibitem{IceCube:2018dna}
The IceCube Collaboration, Fermi-LAT, MAGIC {et al}., 
\MYhref[journalLinks]{https://doi.org/10.1126/science.aat1378}{Science {\bf 361}, eaat1378 (2018)};
IceCube Collaboration: M.~G.~Aartsen {et al}., 
\MYhref[journalLinks]{https://doi.org/10.1126/science.aat2890}{Science {\bf 361}, 147 (2018)}.

\bibitem{Li:2020uef}
H.~Li and B.-Q.~Ma, 
\MYhref[journalLinks]{https://doi.org/10.1016/j.scib.2019.11.024}{Sci. Bull. {\bf 65}, 262 (2020)}.

\bibitem{Li:2021tcv}
C.~Li and B.-Q.~Ma, 
\MYhref[journalLinks]{https://doi.org/10.1103/PhysRevD.104.063012}{Phys. Rev. D {\bf 104}, 063012 (2021)};
\MYhref[journalLinks]{https://doi.org/10.1016/j.scib.2021.07.030}{Sci. Bull. {\bf 66}, 2254 (2021)}.

\bibitem{Ellis:2003sfd}
J.~R.~Ellis, N.~E.~Mavromatos and A.~S.~Sakharov, 
\MYhref[journalLinks]{https://doi.org/10.1016/j.astropartphys.2003.12.001}{Astropart. Phys. {\bf 20}, 669 (2004)};
J.~Ellis, N.~E.~Mavromatos, D.~V.~Nanopoulos and A.~S.~Sakharov, 
\MYhref[journalLinks]{https://doi.org/10.1038/nature02481}{Nature~(London) {\bf 428}, 386 (2004)};
\MYhref[journalLinks]{https://doi.org/10.1142/S0217751X04019780}{Int. J. Mod. Phys. A {\bf 19}, 4413 (2004)}.

\bibitem{Li:2022ugz}
C.~Li and B.-Q.~Ma, 
\MYhref[journalLinks]{https://doi.org/10.1016/j.physletb.2022.137034}{Phys. Lett. B {\bf 829}, 137034 (2022)}.

\bibitem{He:2022jdl}
P.~He and B.-Q.~Ma, 
\MYhref[journalLinks]{https://doi.org/10.1016/j.physletb.2022.137536}{Phys. Lett. B {\bf 835}, 137536 (2022)}.

\bibitem{Crivellin:2020oov}
A.~Crivellin, F.~Kirk and M.~Schreck, 
\MYhref[journalLinks]{https://doi.org/10.1007/JHEP04(2021)082}{JHEP {04} (2021) 082}.

\bibitem{Cohen:2011hx}
A.~G.~Cohen and S.~L.~Glashow, 
\MYhref[journalLinks]{https://doi.org/10.1103/PhysRevLett.107.181803}{Phys. Rev. Lett.  {\bf 107}, 181803 (2011)}.

\bibitem{Borriello:2013ala}
E.~Borriello, S.~Chakraborty, A.~Mirizzi and P.~D.~Serpico, 
\MYhref[journalLinks]{https://doi.org/10.1103/PhysRevD.87.116009}{Phys. Rev. D {\bf 87}, 116009 (2013)}.

\bibitem{Stecker:2013jfa}
F.~W.~Stecker, 
\MYhref[journalLinks]{https://doi.org/10.1016/j.astropartphys.2014.02.007}{Astropart. Phys. {\bf 56}, 16 (2014)}.

\bibitem{Diaz:2013wia}
J.~S.~Diaz, V.~A.~Kosteleck\'y and M.~Mewes, 
\MYhref[journalLinks]{https://doi.org/10.1103/PhysRevD.89.043005}{Phys. Rev. D {\bf 89}, 043005 (2014)}.

\bibitem{Wang:2020tej}
K.~Wang, S.-Q.~Xi, L.~Shao, R.-Y.~Liu, Z.~Li and Z.-K.~Zhang, 
\MYhref[journalLinks]{https://doi.org/10.1103/PhysRevD.102.063027}{Phys. Rev. D {\bf 102}, 063027 (2020)}.

\bibitem{Ellis:2000sf}
J.~Ellis, N.~E.~Mavromatos and D.~V.~Nanopoulos, 
\MYhref[journalLinks]{https://doi.org/10.1103/PhysRevD.63.124025}{Phys. Rev. D {\bf 63}, 124025 (2001)}.

\bibitem{Stecker:2014oxa}
F.~W.~Stecker, S.~T.~Scully, S.~Liberati and D.~Mattingly, 
\MYhref[journalLinks]{https://doi.org/10.1103/PhysRevD.91.045009}{Phys. Rev. D {\bf 91}, 045009 (2015)}.

\bibitem{IceCube:2021rpz}
IceCube Collaboration: M.~G.~Aartsen {et al}., 
\MYhref[journalLinks]{https://doi.org/10.1038/s41586-021-03256-1}{Nature~(London) {\bf 591}, 220 (2021)}~[{\it Erratum} \MYhref[journalLinks]{https://doi.org/10.1038/s41586-021-03450-1}{{\bf 592}, E11 (2021)}].

\bibitem{Glashow:1960zz}
S. L. Glashow, 
\MYhref[journalLinks]{https://doi.org/10.1103/PhysRev.118.316}{Phys. Rev. {\bf 118}, 316 (1960)}.

\bibitem{IceCube:2014noc}
See, e.g., Super-Kamiokande Collaboration: K.~Abe {et al}., 
\MYhref[journalLinks]{https://doi.org/10.1103/PhysRevD.91.052003}{Phys. Rev. D {\bf 91}, 052003 (2015)};
IceCube Collaboration: M.~G.~Aartsen {et al}., 
\MYhref[journalLinks]{https://doi.org/10.1038/s41567-018-0172-2}{Nat. Phys. {\bf 14}, 961 (2018)};
R.~Abbasi {et al}., 
\MYhref[journalLinks]{https://doi.org/10.1038/s41567-022-01762-1}{Nat. Phys. {\bf 18}, 1287 (2022)}.

\bibitem{Lisi:2000ndc}
See, for instance, E.~Lisi, A.~Marrone and D.~Montanino, 
\MYhref[journalLinks]{https://doi.org/10.1103/PhysRevLett.85.1166}{Phys. Rev. Lett. {\bf 85}, 1166 (2000)};
G.~L.~Fogli, E.~Lisi, A.~Marrone, D.~Montanino and A.~Palazzo, 
\MYhref[journalLinks]{https://doi.org/10.1103/PhysRevD.76.033006}{Phys. Rev. D {\bf 76}, 033006 (2007)};
IceCube Collaboration: R.~Abbasi {et al}., 
\MYhref[journalLinks]{https://doi.org/10.1103/PhysRevD.79.102005}{Phys. Rev. D {\bf 79}, 102005 (2009)}.

\end{thebibliography}
\end{document}